\documentclass[prb, superscriptaddress, showpacs]{revtex4}
\usepackage{epsfig}
\usepackage{graphicx}
\usepackage{amsmath}
\usepackage{bm}
\usepackage{amssymb}
\usepackage{color}

\newcommand{\be}{\begin{equation}}
\newcommand{\ee}{\end{equation}}
\newcommand{\ba}{\begin{eqnarray}}
\newcommand{\ea}{\end{eqnarray}}

\definecolor{purple}{rgb}{0.8,0,0.6}

\makeatletter
\newcommand{\vast}{\bBigg@{2}}
\newcommand{\Vast}{\bBigg@{3}}
\makeatother

\begin{document}

\title{Supercriticality of novel type induced by electric dipole in gapped graphene}
\date{\today}

\author{E. V. Gorbar}
\affiliation{Department of Physics, Taras Shevchenko National University of Kiev, Kiev, 03680, Ukraine}
\affiliation{Bogolyubov Institute for Theoretical Physics, Kiev, 03680, Ukraine}

\author{V. P. Gusynin}
\affiliation{Bogolyubov Institute for Theoretical Physics, Kiev, 03680, Ukraine}

\author{O. O. Sobol}
\affiliation{Department of Physics, Taras Shevchenko National University of Kiev, Kiev, 03680, Ukraine}

\begin{abstract}
We reveal a new type of supercritical behavior in gapped graphene with two oppositely charged
impurities by studying the two-dimensional Dirac equation for quasiparticles with the Coulomb 
potential regularized at small distances accounting the lattice effects. By utilizing the variational Galerkin--Kantorovich method, we show that 
for supercritical electric dipole the wave function of the electron bound state changes its localization from the negatively charged impurity to 
the positively charged one as the distance between the impurities changes. Such a migration of the wave function corresponds to the electron and 
hole spontaneously created from the vacuum in bound states screening 
the positively and negatively charged impurities of the supercritical electric dipole, respectively. 
We generalize our results to a particle-hole asymmetric case, where the charges of impurities differ 
in signs and absolute values and demonstrate that the necessary energetic condition for the 
supercriticality of novel type to occur is that the energy levels of single positively and negatively
charged impurities traverse together the energy distance separating the upper and lower continua. The robustness of the supercriticality of 
novel type is confirmed by the study of an exactly solvable 1D
problem of the Dirac equation with the square well and barrier potential modeling an electric dipole potential.
\end{abstract}
\maketitle

\section{Introduction}

It is well known that the Dirac Hamiltonian for the  electron in the Coulomb field of a point charge
$Ze$ is not self-adjoint for $Z > 137$ and the energy of the lowest $1S_{1/2}$ bound state $E=m\sqrt{1-Z^2\alpha^2}$ becomes imaginary 
testifying the fall into the center (atomic collapse)
phenomenon \cite{Pomeranchuk,Zeldovich,Greiner}. Pomeranchuk and Smorodinsky showed \cite{Pomeranchuk} that this problem disappears if a finite 
size of nuclei is taken into account. Then a physically acceptable solution exists for larger values of the charge and its energy is real.
Still the lowest energy electron bound state dives into the lower continuum for $Z \gtrsim 170$ leading
to the spontaneous creation of electron-positron pairs with the electrons screening the positively 
charged nucleus and the positrons emitted to infinity \cite{Zeldovich,Greiner}. Since supercritically charged nuclei are not encountered in 
nature, this phenomenon was never observed in quantum electrodynamics.

It is well known that quasiparticles in graphene are described by the two-dimensional Dirac equation and their interaction with the 
electromagnetic field is characterized by the large effective coupling constant
$\alpha_g=e^2/(\hbar v_F)\approx 2.2$, $v_F \approx c/300$ being the Fermi velocity ($c$ is the velocity 
of light). Therefore, the value of the critical charge in graphene dramatically decreases and equals $Z_c \approx 1/2$
\cite{Pereira,Shytov,Novikov}. We would like to mention also that the supercritical Coulomb center instability is closely related to the 
excitonic instability in graphene in the strong coupling regime
$\alpha_g>\alpha_c\sim1$ (see Refs.[\onlinecite{excitonic-instability,Fertig,Guinea}]) and possible gap opening, which may transform graphene 
into an insulator \cite{metal-insulator,GGG2010,MS-phase-transition,Gonzalez}.

One would think that the supercritical Coulomb center instability due to the large value of the coupling constant should be easily observed in 
graphene. However, it is difficult in practice to produce highly charged impurities. In addition, the external charge in a realistic 
experimental set-up should be smeared over a finite region of the graphene plane because, otherwise, the Dirac equation is no longer applicable 
and other nearest $\sigma$-bands should be included in the analysis \cite{Novikov}. Therefore, the experimental observation of the supercritical 
instability in graphene was not demonstrated until recently. 
A clever means to solve this problem was recently proposed and realized experimentally in Ref.[\onlinecite{Wang}]. By creating artificial nuclei 
in a certain region of graphene fabricated through the deposition of charged calcium dimers on graphene with the tip of a scanning tunneling 
microscope, the supercritical regime was reached and the resonances corresponding to the atomic collapse states were observed.

By making use of the density functional theory and an improved Huckel model, the supercritical instability for ${\rm C}a$ dimers on graphene was 
theoretically studied in Ref.[\onlinecite{Kirczenow}]. An ``atomic-collapse'' state in graphene was found for fewer absorbed ${\rm C}a$ dimers
than in the experiment, possibly due to the different spacing between dimers and the dielectric screening by a boron nitride 
substrate. In the continuum model, the study of the supercritical instability of one Coulomb center in gapped graphene was extended by us 
\cite{two-centers} to the case of the simplest cluster of two equally charged impurities when the charges
of impurities are subcritical, whereas their total charge exceeds a critical one. We determined the critical distance between the impurities 
separating the supercritical and subcritical regimes as a function of
charges of impurities and a gap.

An interesting electric dipole problem in gapped graphene with two oppositely charged impurities was recently considered in 
Refs.~[\onlinecite{Egger,Matrasulov}] (the 3D Dirac equation with the electric 
dipole potential was also studied some time ago in Ref.~[\onlinecite{Matveev}]). It was shown that the 
point electric dipole potential accomodates towers of infinitely many bound states exhibiting a universal Efimov-like scaling hierarchy and at 
least one infinite tower of bound states exists for an arbitrary dipole strength. Notice that the Schr$\ddot{o}$dinger equation in two 
dimensions for the electron in the field of an electric dipole also admits a bound state for any dipole strength \cite{Connolly} unlike the 
three-dimensional case where a bound state exists only when the dipole moment exceeds a certain critical value (see, e.g., a discussion 
including historical one in Ref.~[\onlinecite{Turner}]). By combining  analytical and numerical methods, Ref.~[\onlinecite{Egger}] found that 
the bound states do not dive into the lower continuum because the positive and negative energy levels first approach each other and then go 
away.
Actually this behavior is typical for an avoided crossing \cite{Wigner}, which forbids level crossing for two states with the same quantum 
numbers. Since the bound states do not dive into the lower continuum, the authors of Ref.[\onlinecite{Egger}] concluded that supercriticality is 
unlikely to occur in the electric dipole problem in graphene.

We reconsidered the problem of supercriticality in our recent paper [\onlinecite{supercriticality}] for the case of two oppositely charged 
impurities situated at finite distance (finite electric dipole). By using the linear combination of atomic orbitals (LCAO) and variational 
Galerkin--Kantorovich (GK) methods, we showed that for sufficiently large charges of impurities the wave function of the highest energy occupied 
bound state changes its localization from the negatively charged impurity to the positive one as the distance between the impurities changes 
(both methods gave similar results). The necessary condition for the instability to occur is the crossing of the electron energy levels in the 
field of single positively and negatively charged impurities. This migration of the electron wave function of the supercritical electric dipole 
is a generalization of the familiar phenomenon of the atomic collapse of a single charged impurity with holes emitted to infinity to 
the case where both electrons and holes are spontaneously created from the vacuum in bound states with two oppositely charged impurities thus 
partially screening them.

In this paper, we extend by using the GK method the study of the electric dipole problem in gapped graphene performed in
Ref.[\onlinecite{supercriticality}]. We apply this method also to the more general case of two oppositely charged impurities whose charges are 
not equal by modulus. In this case, we find that the wave function of the highest energy occupied bound state changes its localization
only if the bound state levels in the corresponding one Coulomb center problems traverse together the energy distance $2\Delta$ ($\Delta$ is 
a gap). Since the LCAO and variational GK methods are approximate ones because in practice one can take into consideration only a finite number 
of trial functions, it is important to check the robustness and validity of the supercriticality of novel type connected with the migration of 
the wave function of the electron bound state. For this, we study in this paper an exactly solvable 1D model of the Dirac equation with the 
square well and barrier potential modeling an electric dipole potential. In addition, this model allows one to consider large 
constituent charges in a dipole when oscillations of energy levels and connected with them the changes of the localization of the electron wave 
function take place.

The paper is organized as follows. The supercritical instability in an exactly solvable 1D problem is investigated in Sec.\ref{A}. In Sec.
\ref{section-equation}, we consider the symmetry properties of the 
Dirac equation for quasiparticles in the field of an electric dipole in graphene. In Sec.\ref{variational}, we solve the Dirac equation and 
study the supercritical instability in graphene in the electric
dipole potential by means of the numerical variational Galerkin--Kantorovich method. An asymmetric problem of two impurities with different by 
modulus charges of opposite sign is considered in Sec.\ref{asymmetric}. The results are discussed in Conclusion. Oscillations of energy levels  
in the exactly
solvable 1D problem are considered for sufficiently large strength of an electric-dipole-like potential
in Appendix \ref{2nd-in-1Dmodel}. The system of differential equations in the variational Galerkin--Kantorovich method is written down in 
Appendix \ref{B}.

\section{One-dimensional model with an electric-dipole-like potential}
\label{A}

\begin{figure}[ht]
  \centering
  \includegraphics[scale=0.45]{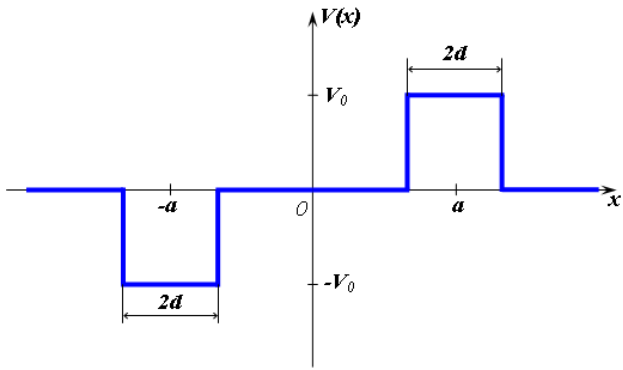}
  \caption{(Color online) An ``electric-dipole-like'' potential of the 1D model.}
  \label{fig1-toy-model}
\end{figure}

In this section, we consider the supercriticality of novel type and the effect of relocalization of the wave function of the highest energy 
occupied bound electron state in an exactly solvable one-dimensional model of the Dirac equation with the square potential well and 
barrier modeling an electric dipole potential \cite{footnote}. This model can be applied also for the description of edge states in graphene
in the presence of a dipole on a boundary. The Hamiltonian of this problem reads
\begin{equation}
H(\Delta,V_0)=-i\hbar v_{F}\sigma_{x}\partial_{x}+\Delta \sigma_{z}+V(x),
\label{hamiltonian-1D}
\end{equation}
where $\sigma_x$ and $\sigma_z$ are the Pauli matrices, $\Delta$ is a gap, and an ``electric-dipole-like'' potential $V(x)$ is defined by the 
equation
\begin{equation}
V(x)=-V_0\theta\left(d-|x+a|\right)+V_0\theta\left(d-|x-a|\right)
\label{dipole-1D}
\end{equation}
and is plotted in Fig.\ref{fig1-toy-model}. Here $2d$ is the width of the well and barrier, and $2a$
is the distance between their centers. In what follows, we will use dimensionless quantities $\epsilon=E/\Delta$, $v_{0}=V_{0}/\Delta$, 
$x\rightarrow xR_{\Delta}=x\hbar v_{F}/\Delta$, $a\rightarrow a\hbar v_{F}/\Delta$, and $d\rightarrow d\hbar v_{F}/\Delta$ (note that
$\hbar v_{F}/\Delta$ is the Compton wavelength).
Hamiltonian (\ref{hamiltonian-1D}) with potential (\ref{dipole-1D}) has a particle-hole symmetry expressed by
$\Omega H(\Delta,V_0)\Omega^+=-H(\Delta,V_0)$, where the unitary operator $\Omega=\sigma_x{\cal R}_x$, with the operator ${\cal R}_x$ of 
reflection $x\rightarrow -x$, satisfies $\Omega^2=1$. It follows then that all solutions of the Dirac equation come in pairs with
$\pm \epsilon$.

\begin{figure}[ht]
  \centering
  \includegraphics[scale=0.4]{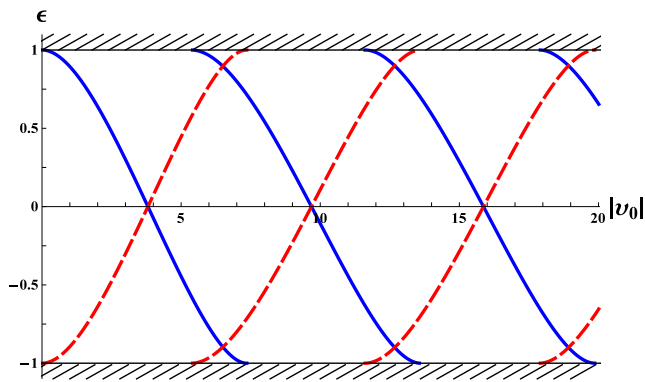}
  \caption{(Color online) The dependence of energy of bound states in the potential well/barrier
  on the strength of the potential $|v_0|$ at the fixed value $d=0.25$: blue solid lines correspond to the potential well ($v_0>0$) and red 
  dashed lines to the barrier ($v_0<0$).}
  \label{1well-energy}
\end{figure}

We will see below that the behavior of the energy levels and wave functions in the model with the ``electric-dipole-like'' potential depends on 
the sign of energy of the electron bound states in the single potential well and barrier problems. For the potential
$v(x)=-v_0\theta\left(d-|x|\right)$, we plot in Fig.\ref{1well-energy} the dependence of the energy levels of electron bound states on $|v_{0}|$ 
for the potential well and barrier at the fixed value $d=0.25$. The energy levels in the potential well with $v_0>0$ 
are plotted by solid blue lines and the corresponding levels for the potential barrier of the same strength but opposite sign potential $v_0<0$ 
are shown by red dashed lines. As is seen, bound states appear for an arbitrary small $|v_0|$. As $|v_0|$ increases, the energy levels cross 
$\epsilon = 0$ and then enter the continua for certain critical strengths (for example, $|v_0|\approx 7.5$ for the first
levels). The energies of the first bound state levels change their sign at $|v_{0}|\approx 4$.

We will solve the Dirac equation with the electric-dipole-like potential in each of five regions defined 
by Eq.(\ref{dipole-1D}) and then match solutions at the points where the potential jumps by using the condition of continuity of each component 
of the spinor. Then, for the component $\phi$ of the two-component spinor $|\Psi\rangle=\left(\phi,\ \chi\right)^{T}$, we find the equation
\begin{equation}
\label{shroedinger_equation}
\phi''+\left((v-\epsilon)^2-1\right)\phi=0,
\end{equation}
and $\chi$ is related to $\phi$ through $\chi=i\partial_{x}\phi/(v-\epsilon-1)$. We have the following solutions:

1)\, $x<-a-d$
\begin{eqnarray}
\phi_{I}(x)&=&C_{1}e^{\kappa_{0}x},\nonumber\\
\chi_{I}(x)&=&-iC_{1}\frac{\kappa_{0}}{\epsilon+1}e^{\kappa_{0}x};
\end{eqnarray}

2)\, $-a-d< x < -a+d$
\begin{eqnarray}
\phi_{II}(x)&=&C_{2}\sin(k_{1}x)+C_{3}\cos(k_{1}x),\nonumber\\
\chi_{II}(x)&=&-i\frac{k_{1}}{v_{0}+\epsilon+1}\left(C_{2}\cos(k_{1}x)-C_{3}\sin(k_{1}x)\right);
\end{eqnarray}

3)\, $-a+d< x < a-d$
\begin{eqnarray}
\phi_{III}(x)&=&C_{4}e^{\kappa_{0}x}+C_{5}e^{-\kappa_{0}x},\nonumber\\
\chi_{III}(x)&=&-i\frac{\kappa_{0}}{\epsilon+1}\left(C_{4}e^{\kappa_{0}x}-C_{5}e^{-\kappa_{0}x}\right);
\end{eqnarray}

4)\, $a-d< x < a+d$
\begin{eqnarray}
\phi_{IV}(x)&=&C_{6}\sin(k_{2}x)+C_{7}\cos(k_{2}x),\nonumber\\
\chi_{IV}(x)&=&i\frac{k_{2}}{v_{0}-\epsilon-1}\left(C_{6}\cos(k_{2}x)-C_{7}\sin(k_{2}x)\right);
\end{eqnarray}

5)\, $x>a+d$
\begin{eqnarray}
\phi_{V}(x)&=&C_{8}e^{-\kappa_{0}x},\nonumber\\
\chi_{V}(x)&=&iC_{8}\frac{\kappa_{0}}{\epsilon+1}e^{-\kappa_{0}x}.
\end{eqnarray}

Here we defined $\kappa_{0}=\sqrt{1-\epsilon^2}$, $k_{1}=\sqrt{(v_{0}+\epsilon)^2-1}$, and $k_{2}=\sqrt{(v_{0}-\epsilon)^2-1}$.
By matching the solutions
\begin{equation}
\phi(x_{i}+0)=\phi(x_{i}-0),\ \ \ \ \chi(x_{i}+0)=\chi(x_{i}-0), \ \ \ i=\overline{1,4}
\end{equation}
at the four points
\begin{equation}
x_{1}=-(a+d),\ \ \ \ x_{2}=-(a-d), \ \ \ \ x_{3}=a-d, \ \ \ \ x_{4}=a+d,
\end{equation}
we find the following system of equations for coefficients $C_{k}$:
\begin{equation}
A_{ik}C_k=0,
\label{system-of-eqs}
\end{equation}
where the coefficients $A_{ik}$ are given by Eq.(\ref{Aij-coeff}) in Appendix A. The secular equation 
$\det A=0$ determines the bound state energy levels of the problem. For $a<d$, the energy levels are obtained by the interchange 
$a\leftrightarrow d$. The coefficients $C_k$ determined from Eq.(\ref{system-of-eqs}) together with a normalization condition specify completely 
the wave functions.

\begin{figure}[ht]
  \centering
  \includegraphics[scale=0.35]{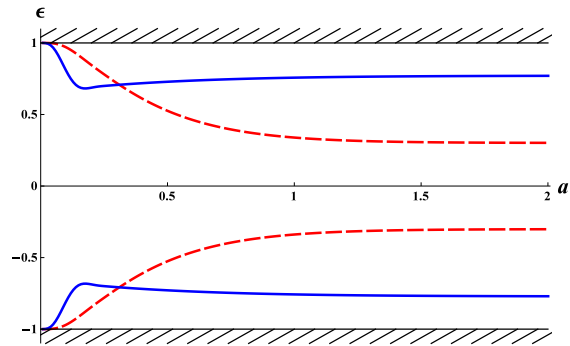}
  \caption{(Color online) The bound state energy levels of the 1D problem with the electric-dipole-like potential for $v_{0}=3$ (red dashed 
  lines) and $v_{0}=6$ (blue thick lines) obtained for $d=0.25$.}
  \label{fig2-toy}
\end{figure}

For $d=0.25$, we plot the bound state energy levels in Fig.\ref{fig2-toy} for two values $v_{0}=3$ and $v_{0}=6$.

\begin{figure}[ht]
  \centering
  \includegraphics[scale=0.42]{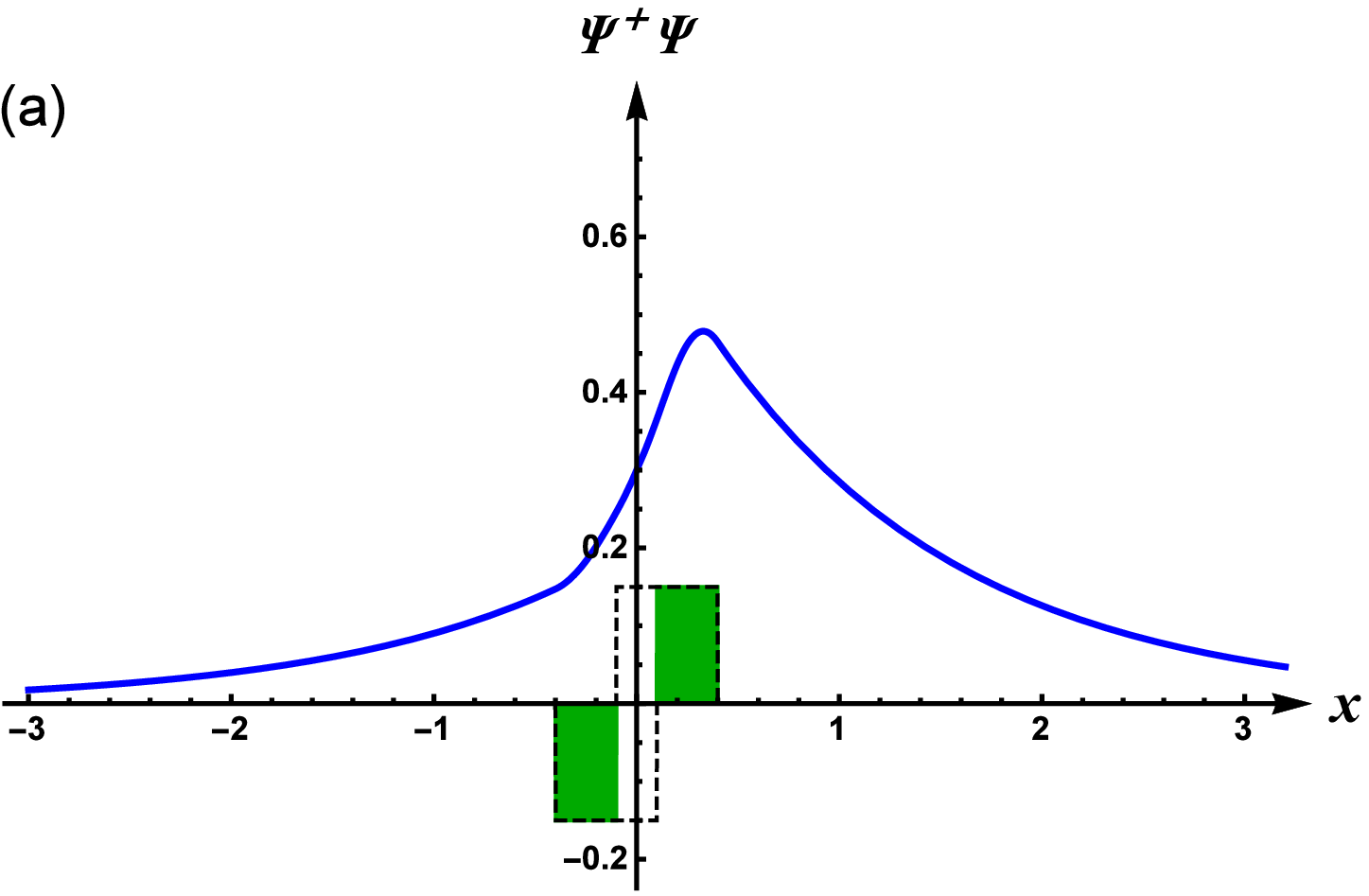}
  \includegraphics[scale=0.4]{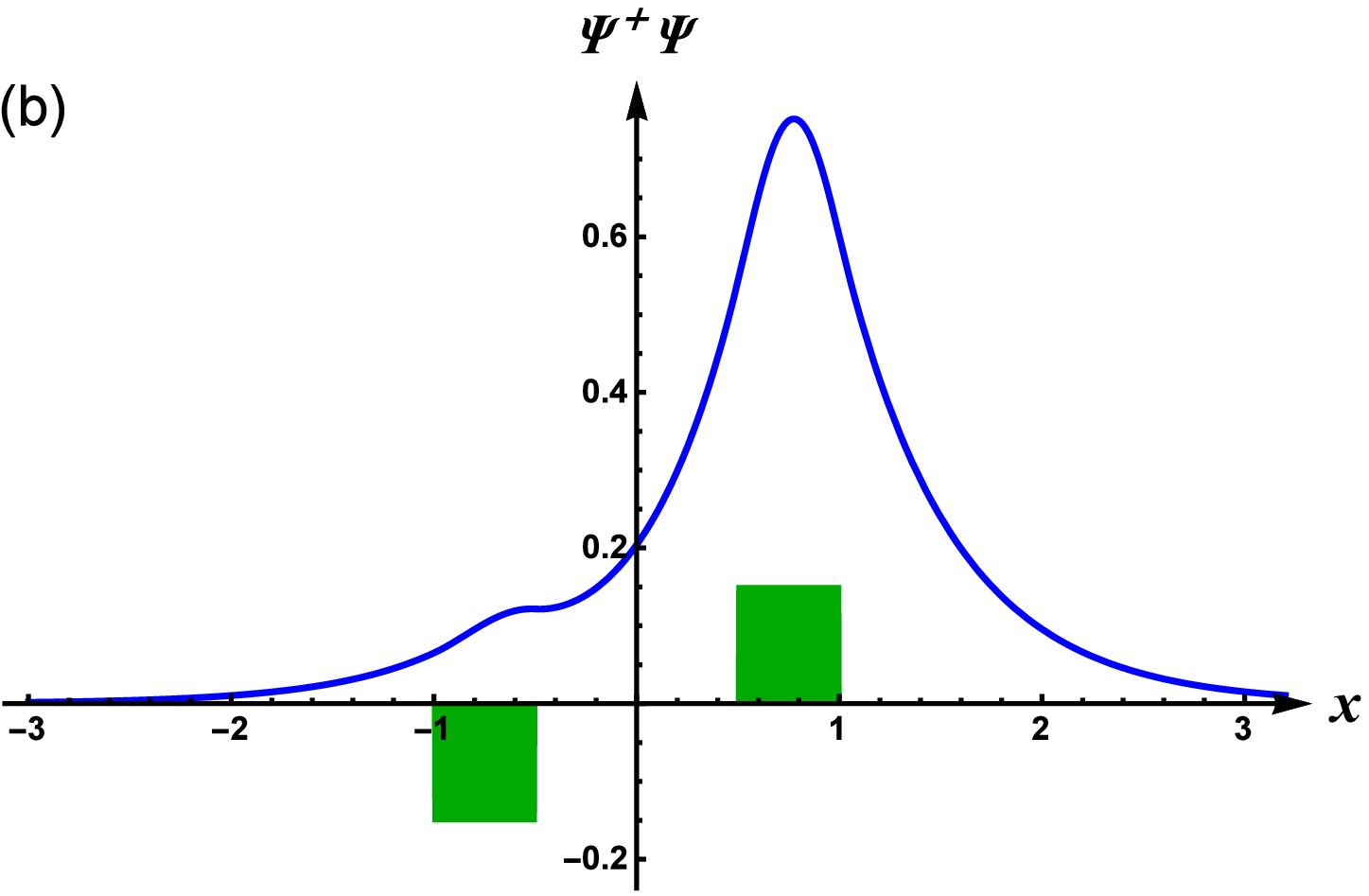}
  \caption{(Color online) The square modulus of the wave function of the negative energy bound state for $v_{0}=3$, $d=0.25$, and two values of 
  the distance between the centers of the well and barrier: (a) $2a=0.3$, and (b) $2a=1.5$. The potentials of the well and barrier are 
  schematically plotted as filled green regions.}
  \label{fig3-toy}
\end{figure}

Fig.\ref{fig2-toy} implies that the energy levels for the smaller value $v_{0}=3$ monotonously depend on the distance between the well and 
barrier. We plot in Fig.\ref{fig3-toy} the square modulus of the wave function of the negative energy bound state for two values
of the distance between the centers of the well and barrier $2a=0.3$ and $2a=1.5$ whose potentials are schematically shown by filled 
green regions. Since the well and barrier overlap for $2a=0.3$, their widths are reduced in this case. For $2a=1.5$, the well and barrier are 
well separated and, obviously, the wave function does not change its localization on the barrier. Note that for $v_0=3$ the energy levels in the 
single potential well problem do not cross the zero energy value (see Fig.~\ref{1well-energy}).

\begin{figure}[h]
  \centering
  \includegraphics[scale=0.42]{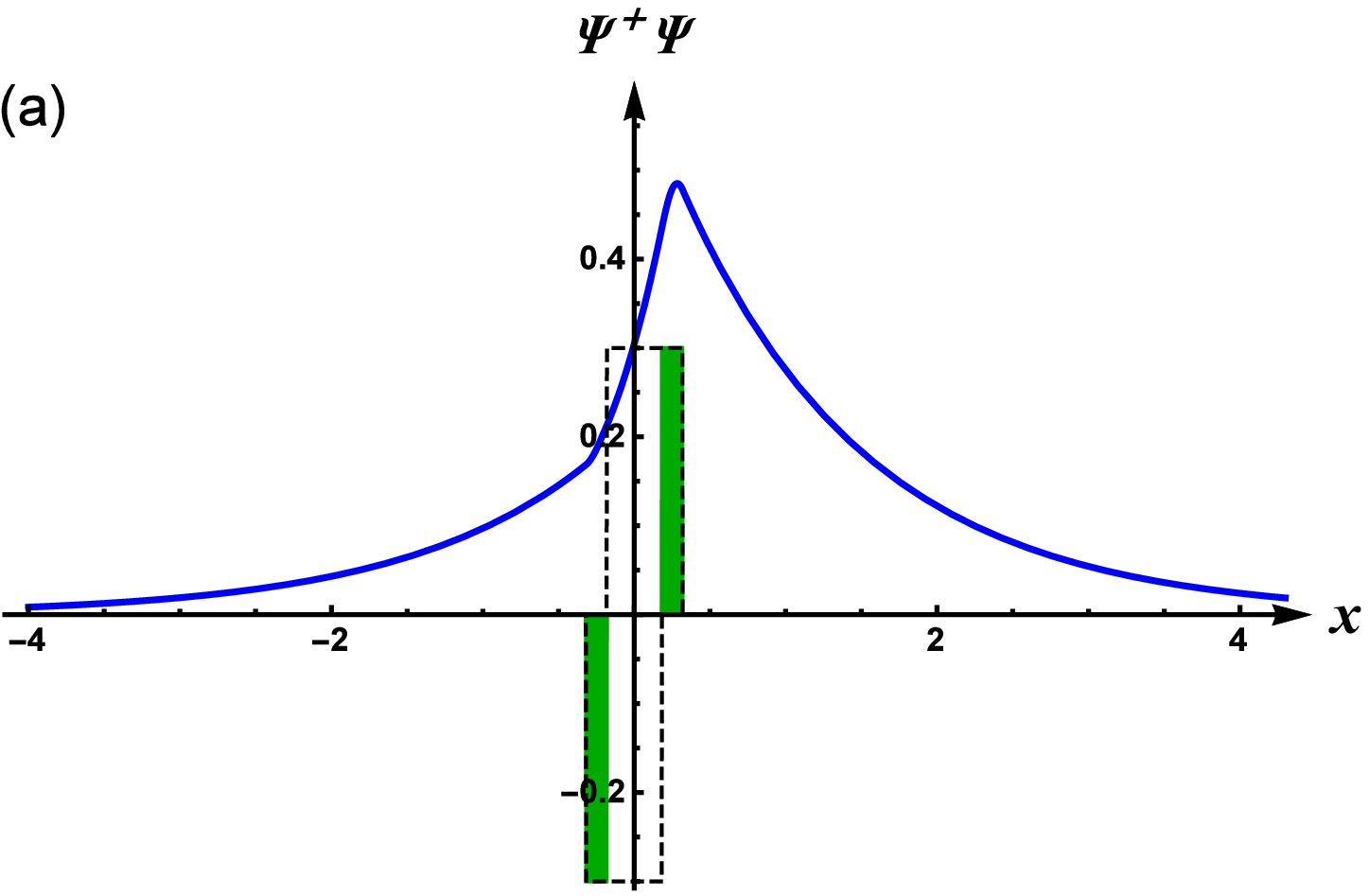}
  \includegraphics[scale=0.42]{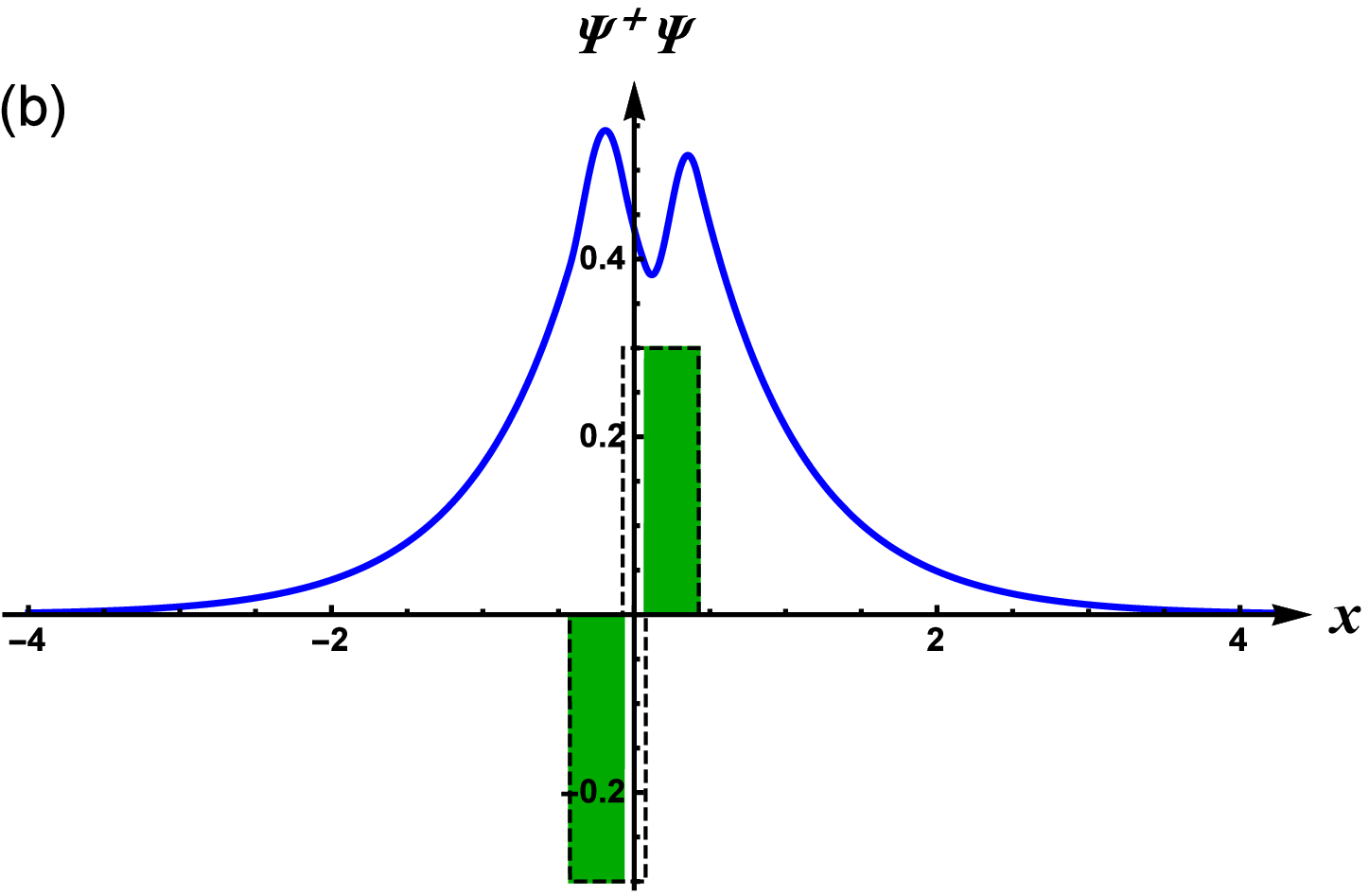}
 \includegraphics[scale=0.42]{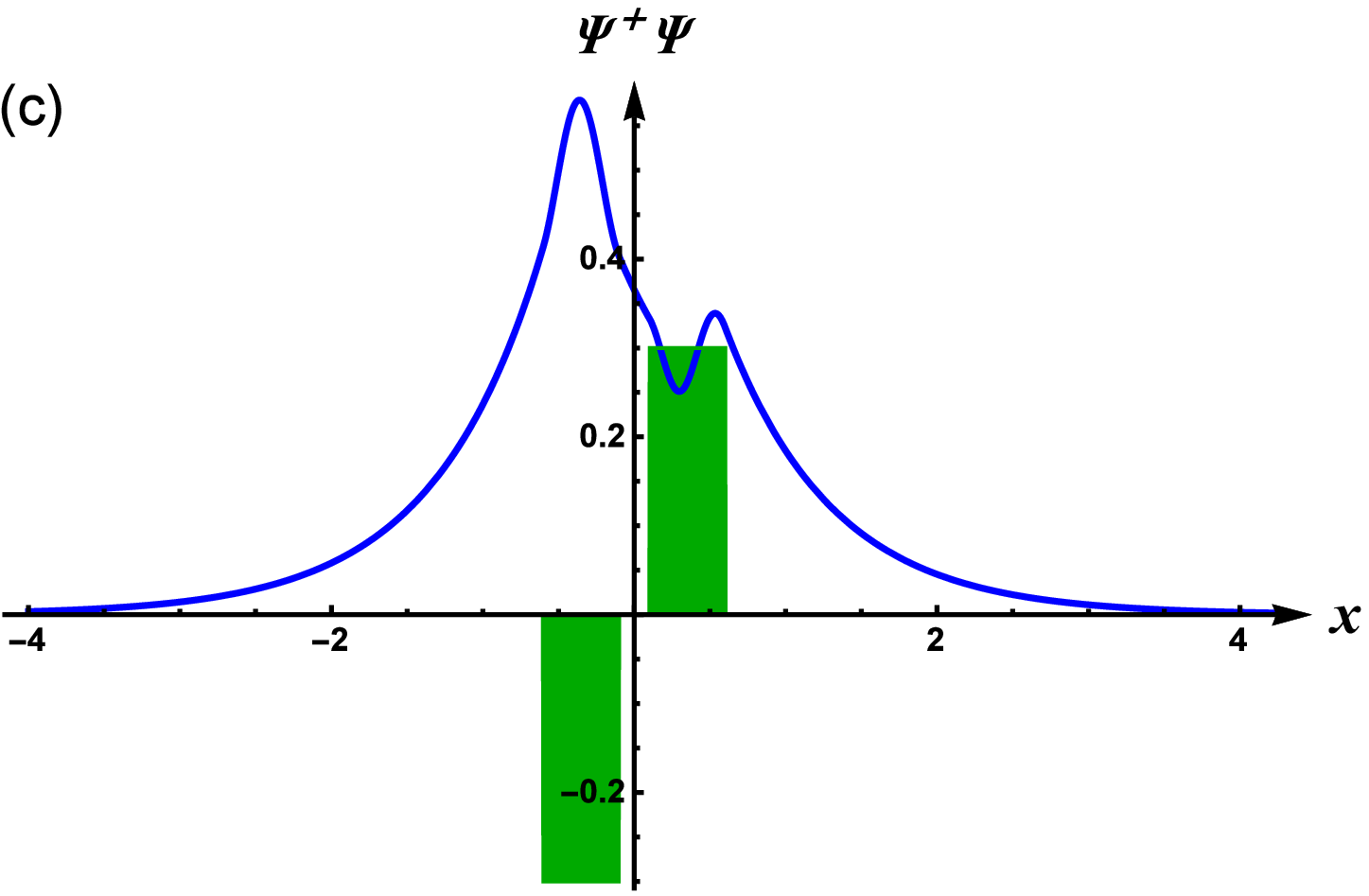}
  \includegraphics[scale=0.42]{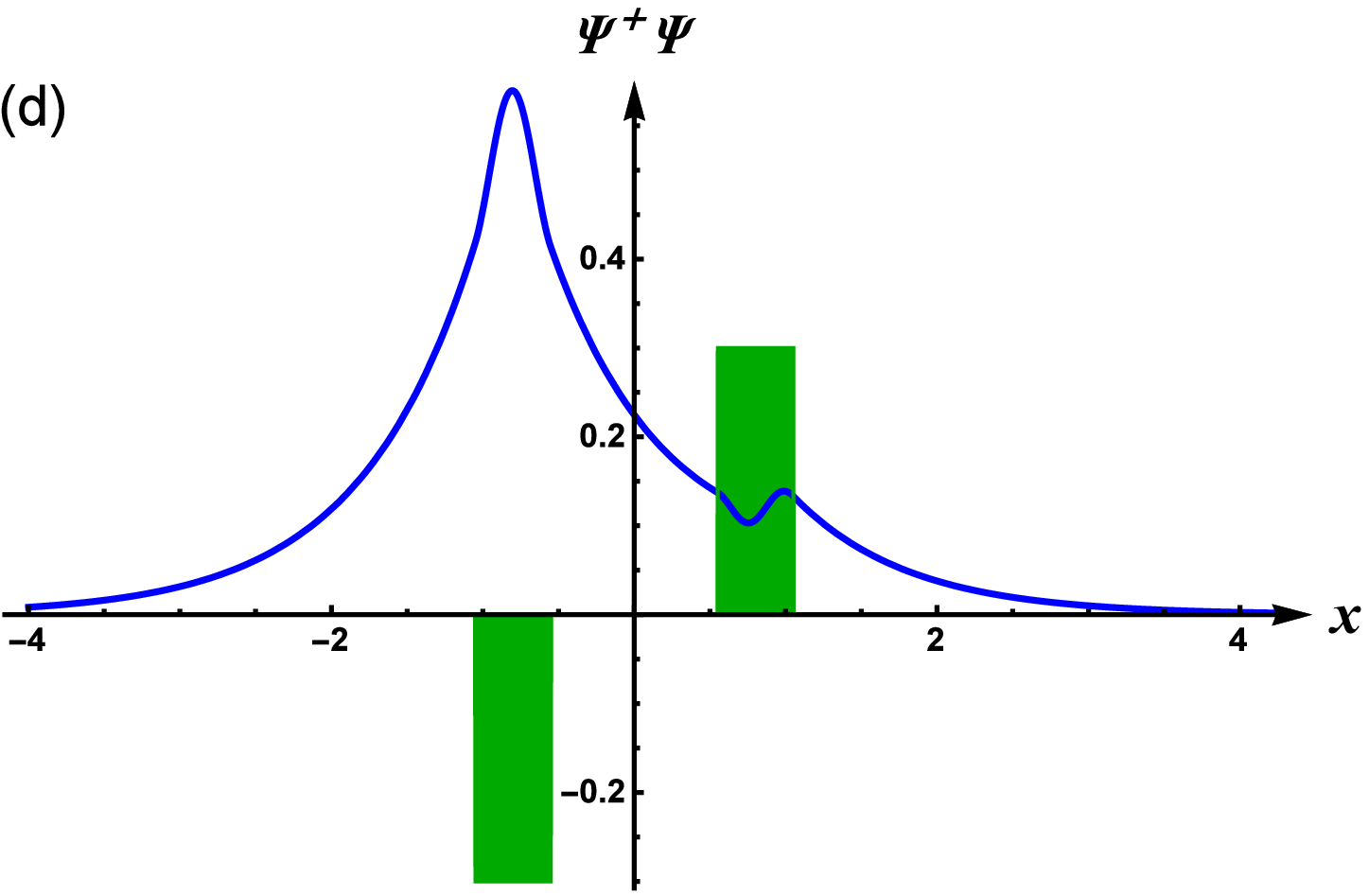}
  \caption{(Color online) The square modulus of the wave function of the negative energy bound state for $v_{0}=6$, $d=0.25$, and four values of 
  the distance between the centers of the well and barrier: (a) $2a=0.14$, (b) $2a=0.35$, (c) $2a=0.7$, and (d) $2a=1.6$. The potentials of the 
  well and barrier are schematically plotted as filled green regions.}
  \label{fig7-toy}
\end{figure}

For the larger value $v_{0}=6$, according to Fig.\ref{fig2-toy}, the level repulsion is observed. We plot
in Fig.\ref{fig7-toy} the square modulus of the wave function of the negative energy bound state for four values of the distance $2a$ between 
the centers of the well and barrier. For $2a=0.14$, the wave function
is localized on the barrier whose width is much reduced due to the strong overlap with the well. For larger value of the distance between the 
centers of the well and barrier $2a=0.35$, the wave function is localized both on the barrier and well. As the distance between the centers 
increases further, the wave function migrates to the well. It is crucial that the case $v_0=6$ corresponds to the negative sign of the energy of 
the first bound state in the potential well problem (see Fig.~\ref{1well-energy}). Note that a similar phenomenon of the change of the 
localization of wave function takes place in the fission of quarkonium consisting of heavy quark and antiquark \cite{Greiner,Vasak}.

For larger values of  $v_0$ when several energy levels cross the zero energy $\epsilon=0$ in the potential well problem in
Fig.\ref{1well-energy}, we observe oscillations in the behavior of energy levels as well as oscillations in the localization of the wave 
function on the well and barrier (see, Appendix \ref{2nd-in-1Dmodel}).

Thus, the exact solutions of the 1D Dirac equation with the electric-dipole-like potential unambiguously confirm the conclusion made in 
Ref.~[\onlinecite{supercriticality}] that the supercritical instability in the presence of both attractive and repulsive potentials is connected 
with the change of the localization of the wave function of the highest occupied electron bound state. 

The local density of states LDOS$(\epsilon,a,x)=\sum_k|\Psi_k(x)|^2\delta(\epsilon-\epsilon(k))$, where $k$ is the set of all
quantum numbers, is a physical quantity that can probe the migration of the wave function and be directly measured in an experiment. Therefore, 
we present the video in Supplemental Material [\onlinecite{animation}], where the LDOS in the 1D model with an electric-dipole-like potential 
considered in this section is plotted as $a$ changes for the energy close to the boundary of the lower continuum $\epsilon=-1.1$
and the three values of $v_0$ such that the energy of the lowest electron bound state in the potential well problem
$\epsilon_0$ equals $0.35$, $-0.085$, and $-0.88$. Obviously, the first case is subcritical because $\epsilon_0$ is positive, therefore, the 
peak of the LDOS remains localized on the barrier as the distance $2a$ between the centers of the well and barrier changes from small to large 
values (note that since the peak is localized on the barrier, it moves to the right as $2a$ increases). The two other cases are supercritical 
and the migration of the peak of the LDOS from the barrier to well is clearly seen as $a$ changes. Since the last case 
$\epsilon_0=-0.88$ is strongly supercritical, the LDOS peak takes much larger values in this case compared to those in the weakly supercritical 
case $\epsilon_0=-0.085$ when the lowest energy bound state level in the square well potential is only slightly below the zero energy. Thus, we 
conclude that measuring the LDOS in the continua makes it possible to demonstrate the supercriticality of novel type and the migration of the 
electron wave function in the electric dipole problem.

\section{Dirac equation}
\label{section-equation}

Let us consider now the supercritical instability in the electric dipole problem in gapped graphene. The Dirac Hamiltonian in $2+1$ dimensions 
which describes the quasiparticle states in the vicinity of the $K_{\pm}$ points of graphene in the field of two oppositely charged impurities 
reads (we set $\hbar=1$)
\begin{equation}
H(\Delta,e)=v_{F}\boldsymbol{\sigma}\boldsymbol{p}+\xi\Delta\sigma_z-eV(\mathbf{r}),
\label{general-Hamiltonian}
\end{equation}
where $-e<0$ is the electron charge, $\boldsymbol{p}=-i\nabla_{\mathbf{r}}$ is the two-dimensional 
canonical momentum, $\sigma_{i}$ are the Pauli matrices, and $\Delta$ is a quasiparticle gap. The latter 
can be opened in graphene in various ways, e.g., due to finite-size effects in graphene nanoribbons \cite{Peres} or by depositing graphene on a 
substrate \cite{Ponomarenko,Song}.
Hamiltonian (\ref{general-Hamiltonian}) acts on two component spinor $\Psi_{\xi s}$ which carries the 
valley ($\xi=\pm)$ and spin ($s=\pm$) indices and we use the standard convention: $\Psi^{T}_{+s}=(\psi_{A},\psi_{B})_{K_{+}s}$, whereas 
$\Psi^{T}_{-s}=(\psi_{B},\psi_{A})_{K_{-}s}$,
and $A,B$ refer to two sublattices of hexagonal graphene lattice. We regularize the Coulomb potential of each impurity by $r_0$, which is of the 
order of the graphene lattice spacing. Then the regularized interaction dipole potential for charged impurities $\pm Q$, $Q=Ze$, situated in the 
$(x,y)$ plane at $(\pm R/2,0$) is given by
\begin{equation}
V\left(\mathbf{r}\right)=\frac{Q}{\kappa}\left(\frac{1}{\sqrt{(x+R/2)^{2}+y^{2}+r_{0}^{2}}}
-\frac{1}{\sqrt{(x-R/2)^{2}+y^{2}+r_{0}^{2}}}\right),
\label{dipole-potential}
\end{equation}
where $\kappa$ is the dielectric constant. For the sake of definiteness, we will consider the electrons
only in the $K_+$ valley (the Dirac equation for the electrons in the $K_-$ valley is obtained by
replacing $\Delta$ with $-\Delta$). Since the interaction potential (\ref{dipole-potential}) does not
depend on spin, we will omit the spin index $s$ in wave functions in what follows. The main difficulty
in solving the Dirac equation for the electron in the electric dipole potential is that variables in
this problem are not separable in any known orthogonal coordinate system. Therefore, we will utilize in
this paper the numerical variational Galerkin--Kantorovich method.

Let us discuss the discrete symmetries of Hamiltonian (\ref{general-Hamiltonian}) at fixed 
valley and spin with potential (\ref{dipole-potential}). The parity transformation changes $\Delta$ 
to $-\Delta$ and also reverses the sign of the potential, or equivalently in the case under consideration, it reverses the charge of the 
particle, $e\rightarrow-e$:
\begin{equation}
U_pH(\Delta,e)U_p^{-1} = H(-\Delta,-e),
\end{equation}
where $U_p=\sigma_y{\cal R}_x$ is a unitary operator with the operator ${\cal R}_x$ of reflection $(x,y)\rightarrow (-x,y)$. Hence the wave 
function $\Psi_p(\mathbf{r})=U_p\Psi(\mathbf{r})=\sigma_y\Psi({\cal R}_x\mathbf{r})$ describes states with
the same energy but the opposite sign of the charge and gap.

The  charge conjugation operator ${U}_c=\sigma_x K$, where $K$ is the complex conjugation, interchanges
the Hamiltonians $H(\Delta,e)$ and $H(\Delta,-e)$
\begin{equation}
{U}_cH(\Delta,e){U}_c^{-1}=-H(\Delta,-e).
\label{charge-conjugation}
\end{equation}
Therefore, if the wave function $\Psi(\mathbf{r})$ is a solution of the stationary Dirac equation
$H(\Delta,e)\Psi(\mathbf{r})=E\Psi(\mathbf{r})$, then the charge conjugated wave function
$\Psi_c(\mathbf{r})={U}_c\Psi(\mathbf{r})=\sigma_x\Psi^*(\mathbf{r})$ is an eigenfunction of $H(\Delta,-e)$ but with the eigenvalue $-E$. The 
time-reversed wave function $\Psi_T(\mathbf{r})={U}_T\Psi(\mathbf{r})=\sigma_y\Psi^*(\mathbf{r})$ satisfies the equation $H(-\Delta,e)
\Psi_T(\mathbf{r})=E\Psi_T(\mathbf{r})$ because
\begin{equation}
{U}_TH(\Delta,e){U}_T^{-1}=H(-\Delta,e).
\end{equation}
This reflects the well known fact that a gap at fixed valley and spin in graphene breaks the time
reversal symmetry. This symmetry is effectively preserved in the charge density wave and quantum spin
Hall states \cite{McDonald}, which are characterized by a gap of opposite sign for the other
valley and spin, respectively.

Hamiltonian (\ref{general-Hamiltonian}) with potential (\ref{dipole-potential}) has an intrinsic particle-hole symmetry expressed by
$\Omega H(\Delta,e)\Omega^+=-H(\Delta,e)$, where the unitary
operator $\Omega=\sigma_x {\cal R}_x$ satisfies $\Omega^2=1$ (note that $\Omega$ is the same operator 
as in the 1D model in Sec.\ref{A}). It follows then that an eigenstate $\Psi_E(x,y)$ with energy $E$ 
has a partner $\Psi_{-E}(x,y)=\Omega\Psi_E(x,y)=\sigma_x\Psi_E(-x,y)$ with energy $-E$, hence, all 
solutions of the Dirac equation come in pairs with $\pm E$. In fact, the operator $\Omega$ is nothing 
else as the combination of three symmetry operations C, P, and T: $\Omega={U}_cU_p{U}_T$. The intrinsic particle-hole symmetry of the electric 
dipole problem will play a prominent role in our analysis below.

The Dirac Hamiltonian with the electric dipole potential commutes also with the operator
${U}=\sigma_{z}K\mathcal{R}_{y}$, where  $\mathcal{R}_{y}$ is the operator of reflection $y \rightarrow -y$. The equality ${U}^{2}=1$ implies
that wave functions are split into two classes ${U}|\Psi_{\lambda}\rangle=\lambda|\Psi_{\lambda}\rangle$, where $\lambda=\pm 1$. Since the 
operator $U$ is antilinear, the function $|\Psi_-\rangle$ related to the function $|\Psi_+\rangle$ by means of the phase transformation
$|\Psi_-\rangle=i|\Psi_+\rangle$ is an eigenfunction with $\lambda=-1$. Therefore, there no need to consider functions $|\Psi_-\rangle$. Hence 
the components of wave functions $\langle\mathbf{r}|\Psi_+\rangle=\psi(\mathbf{r})=(\phi,\chi)^T$ satisfy the following constraint 
conditions consistent with the Dirac equation:
\begin{equation}
\left\{
\begin{array}{l}
\phi^{*}(-y)=\phi(y),\\
\chi^{*}(-y)=-\chi(y).
\end{array}
\right.
\label{components}
\end{equation}
It is convenient to work with dimensionless quantities $h=\frac{H}{\Delta}$ and $\epsilon=\frac{E}{\Delta}$. In addition, we assume in what 
follows that all coordinates and
distances are dimensionless and are defined in units of $R_{\Delta}=\frac{\hbar v_{F}}{\Delta}$.
We introduce also dimensionless coupling constant $\zeta=\frac{eQ}{\hbar v_{F}\kappa}$.
Then the Dirac equation reduces to the following system of two coupled ordinary differential equations
of the first order:
\begin{equation}
\left\{
\begin{array}{l}
-i(\partial_{x}+i\partial_{y})\phi+(v-\epsilon-1)\chi=0,\\
-i(\partial_{x}-i\partial_{y})\chi+(v-\epsilon+1)\phi=0.
\end{array}
\right.
\label{system-equations}
\end{equation}
We will find numerical solutions of Eqs.(\ref{system-equations}) in the next section by using the variational GK method in the class of wave 
functions with $\lambda=1$. Still it is instructive
to begin our analysis with the Dirac equation for the electron in graphene with one positively charged impurity
\begin{equation}
h_{p}\Psi=\epsilon\Psi,
\end{equation}
where
\begin{equation}
h_p=-i(\sigma_{x}\partial_{x}+\sigma_{y}\partial_{y})+\sigma_{z}-\frac{\zeta}{\sqrt{r^{2}+r_{0}^{2}}}.
\end{equation}
The Hamiltonian $h_n$ for the electron in the field of negatively charged impurity is obtained
from the Hamiltonian $h_p$ by the change of the sign of the last term in $h_p$.

The wave function of the state with the total angular momentum $j$ in the polar coordinates
$(r,\ \theta)$ has the form
\begin{equation}
\Psi=\left(
\begin{array}{c}
e^{i(j-1/2)\theta}f(r)\\
-i e^{i(j+1/2)\theta}g(r)
\end{array}
\right).
\end{equation}
For the lowest energy electron bound state with $j=1/2$, the Dirac equation takes the form
\begin{equation}
\left\{
\begin{array}{c}
f'=(1+\epsilon-v(r))g,\\
g'+\frac{g}{r}=(1-\epsilon+v(r))f.\end{array}\right.
\end{equation}
To find the boundary conditions for the functions $f$ and $g$ at the origin, we investigate the
asymptotic behavior at $r=0$. The solution of the approximate equation
\begin{equation}
f''+\frac{f'}{r}+\frac{\zeta^{2}}{r_{0}^{2}}f=0
\end{equation}
regular at the origin is given by $f=J_{0}\left({\zeta r}/{r_{0}}\right)$, where $J_0(z)$ is the Bessel function. Therefore, the numerical 
solution  of the Dirac equation should satisfy the boundary conditions $f(0)=1$ and $g(0)=0$.

\begin{figure}[ht]
  \centering
  \includegraphics[scale=0.35]{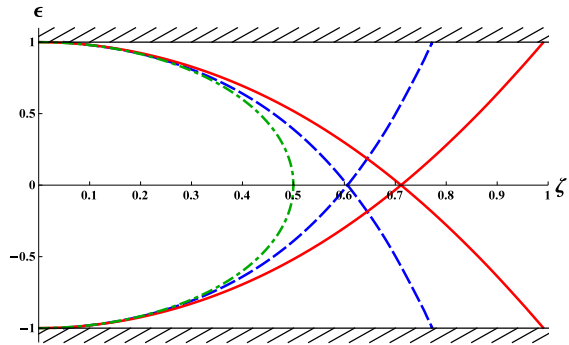}
  \caption{(Color online) The energy of the electron bound state with $j=1/2$ in the regularized
  Coulomb potential  as a function of $\zeta$ for different values of the regularization parameter: $r_{0}=0$ (green dash-dotted lines), 
  $r_{0}=0.01 R_{\Delta}$ (blue dashed lines), $r_{0}=0.05 R_{\Delta}$ (red solid lines). The levels which descend from the upper continuum 
  correspond to the positive charge $+Q$ problem while those which rise from the lower continuum and grow with $\zeta$ correspond to the 
  negative charge $-Q$ problem.}
  \label{fig1-LCAO}
\end{figure}

We determine numerically the energy levels by using the shooting method and requiring that the wave functions decrease at infinity. We normalize 
the wave functions according to the condition
$\int d^{2}r\,\Psi^{+}\Psi=1$. The energy of the lowest (highest)  electron bound state with $j=1/2$
in the regularized Coulomb potential  with the charge $+Q$ ($-Q$) is plotted in Fig.\ref{fig1-LCAO}
for several values of $r_{0}$ as a function of $\zeta$. The levels which descend from the upper continuum correspond to the positively charged 
impurity $+Q$ (these levels agree with the corresponding results in Ref.[\onlinecite{excitonic-instability}], see Fig.4 there), while those 
which rise from the lower continuum and grow with $\zeta$ correspond to the negatively charged impurity $-Q$. These results also
reproduce qualitatively the behavior seen directly at the tight-binding level on a honeycomb lattice [\cite{Kotov}].

For nonregularized Coulomb potential of positive charge, the energy of the lowest bound state is always positive. It reaches the value 
$\epsilon=0$ for $\zeta=1/2$ and becomes purely imaginary for $\zeta>1/2$ (the ``fall into the center phenomenon'' 
\cite{Shytov,Pereira,Novikov}). For regularized Coulomb potential, the energy of the lowest bound state crosses $\epsilon=0$ and approaches the 
negative-energy continuum for a certain critical charge. For example, for $r_{0}=0.05 R_{\Delta}$, the critical charge
$\zeta\approx 1$ corresponds to the lowest electron bound state diving into the lower continuum.

The electron levels in the field of negatively charged center described by the Hamiltonian $h_n$ are obtained by the reflection 
$\epsilon\rightarrow -\epsilon$ because the operator of charge conjugation $U_c=\sigma_xK$ interchanges the $h_p$ and $h_n$ Hamiltonians, 
${U}_{c}h_p{U}^{+}_{c}=-h_n$. Therefore, the energy levels of the Hamiltonians $h_p$ and $h_n$ intersect at $\epsilon=0$. We will see that the 
corresponding critical value of a coupling constant 
$\zeta_c$ plays a crucial role in our analysis of solutions in the electric dipole potential because the behavior of the corresponding energy 
levels dramatically changes depending on whether $\zeta<\zeta_c$ or $\zeta>\zeta_c$ similarly to the analysis of the 1D model in the previous 
section. For the chosen values of the regularization parameter $r_0$ in Fig.\ref{fig1-LCAO}, the critical coupling constant
$\zeta_c\simeq 0.6\, (r_0=0.01R_\Delta)$ and $\zeta_c\simeq 0.7\, (r_0=0.05R_\Delta)$.

Finally, the Hamiltonian for quasiparticles in graphene with two oppositely charged impurities in dimensionless units has the form
\begin{equation}
h=-i(\sigma_{x}\partial_{x}+\sigma_{y}\partial_{y})+\sigma_{z}+\zeta \left(-\frac{1}{\sqrt{r_{p}^{2}+r_{0}^{2}}}
+\frac{1}{\sqrt{r_{n}^{2}+r_{0}^{2}}}\right),
\end{equation}
where $r_{p,n}=\sqrt{(x\pm R/2)^{2}+y^2}$. In order to find eigenstates of this Hamiltonian,
we will apply the variational Galerkin--Kantorovich method in the next section.

\section{Variational method}
\label{variational}

The choice of trial functions is a crucial element in any variational method. We choose the following 
trial functions in the GK method which belong to the class $\lambda=1$ (see the discussion above Eq.(\ref{components})) and satisfy the 
asymptotic at large distance:
\begin{equation}
\label{system2}
\left\{
\begin{array}{l}
\phi(x,y)=e^{-\sqrt{1-\epsilon^{2}}\sqrt{\left(|x|-R/2\right)^{2}+y^{2}+r_{0}^{2}}}
\left\{\sum\limits_{k=0}^{N}f_{2k}(x)y^{2k}+i\sum\limits_{k=1}^{N^{'}}f_{2k-1}(x)y^{2k-1}\right\},\\
\chi(x,y)=-ie^{-\sqrt{1-\epsilon^{2}}\sqrt{\left(|x|-R/2\right)^{2}+y^{2}+r_{0}^{2}}}
\left\{\sum\limits_{k=0}^{N}g_{2k}(x)y^{2k}+i\sum\limits_{k=1}^{N^{'}}g_{2k-1}(x)y^{2k-1}\right\}.
\end{array}
\right.
\end{equation}
According to the Galerkin--Kantorovich method, the wave functions are substituted into the initial equation and their orthogonality to the 
residual with respect to the variable $y$ is required. Thus, we obtain the system of ordinary differential equations
(see Eq.(\ref{variational-method}) in Appendix \ref{B}) for functions $f_k$ and $g_k$.

\begin{figure}[ht]
  \centering
  \includegraphics[scale=0.35]{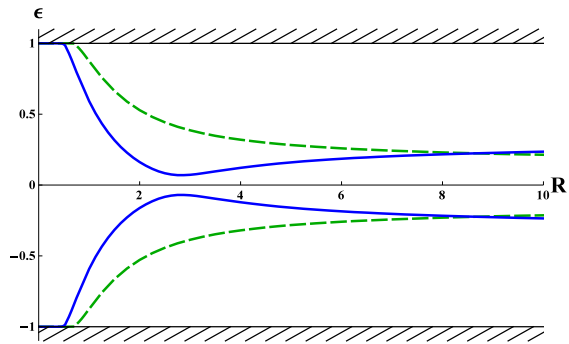}
  \caption{(Color online) The energy levels of bound states obtained in the variational method for $r_{0}=0.05 R_{\Delta}$ and different charges 
  of impurities: $\zeta=0.6$ (green dashed lines) and $\zeta=0.85$ (blue solid lines)}
  \label{sp-variational}
\end{figure}

To determine how successful the GK method is, we first use this method for solving the Dirac equation
for the electron in the field of one Coulomb center whose regularized potential is given by $V(r)=-\frac{\zeta}{\sqrt{r^{2}+r_{0}^{2}}}$. We 
should set $R=0$ in Eqs.(\ref{variational-method}) - (\ref{Q-s}) and take into account that $V_{s}(x)=-\zeta Q_{s}(x)$. The exact spectrum of 
the one Coulomb center problem with $r_{0}=0.05 R_{\Delta}$  plotted in Fig.\ref{fig1-LCAO} (red solid lines) can be compared with the 
approximate spectra determined by the GK method for the various number
of terms in the ansatz: $N=0,N^{'}=0$; $N=1,N^{'}=0$; $N=0,N^{'}=1$; and $N=1,N^{'}=1$.  We checked that
the approximations ``$00$'' and ``$11$'' reproduce the exact spectrum better than the other approximations. Therefore, we use in what follows 
these approximations to determine the bound states for the electron in
the electric dipole potential.

\begin{figure}[ht]
  \centering
  \includegraphics[scale=0.12]{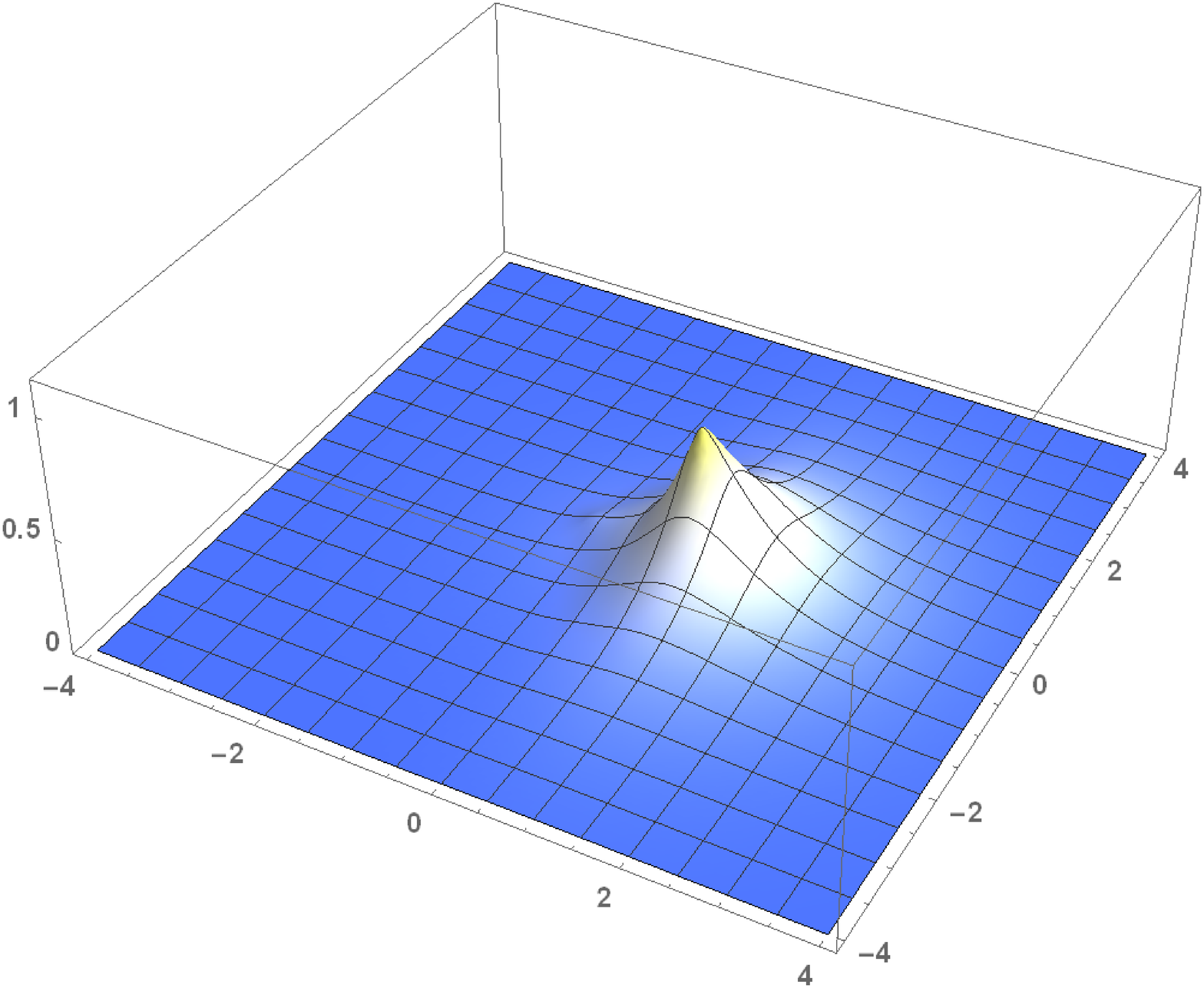}
  \includegraphics[scale=0.12]{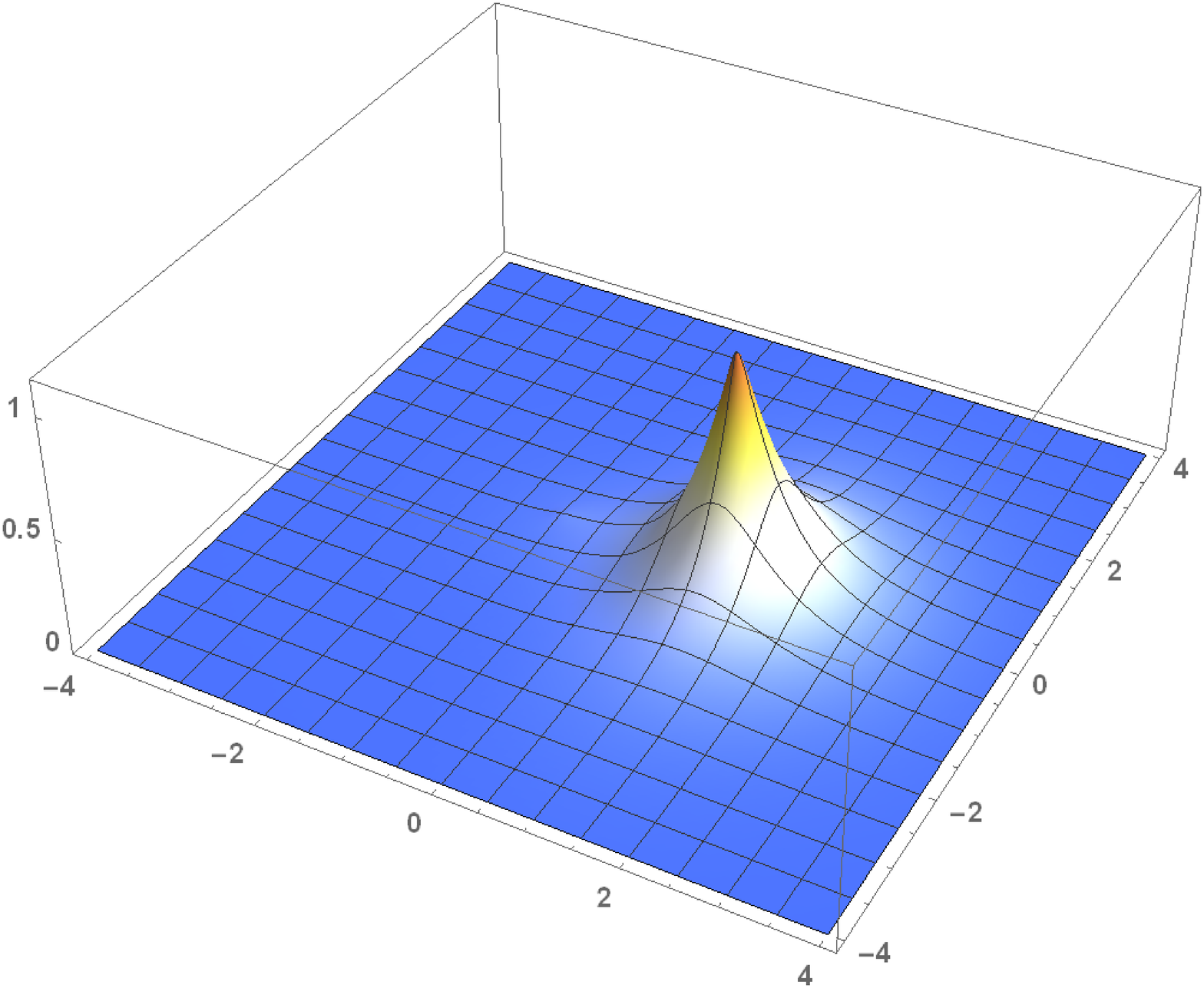}
  \includegraphics[scale=0.12]{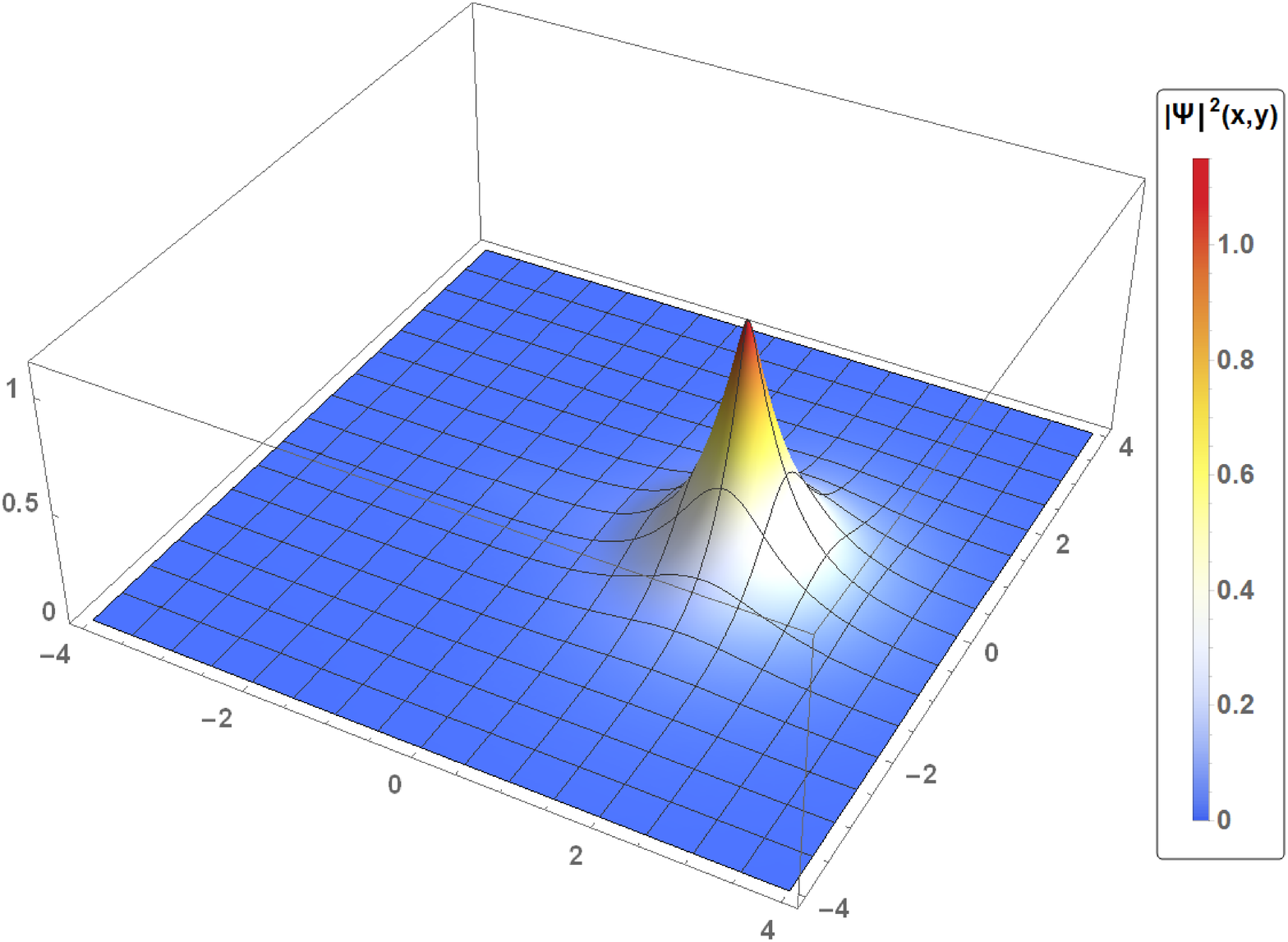}
  \caption{(Color online) The square modulus of the wave function of the lower energy level for $\zeta=0.6$ and various distances between the  
  impurities: $R=1.25$ (left panel), $R=2.0$ (middle panel), and $R=3.0$ (right panel). The wave function doesn't change its localization from 
  the negatively charged
  impurity to the positively charged one as the distance between the impurities increases.}
  \label{Var_no_wf}
\end{figure}
\begin{figure}[ht]
  \centering
  \includegraphics[scale=0.12]{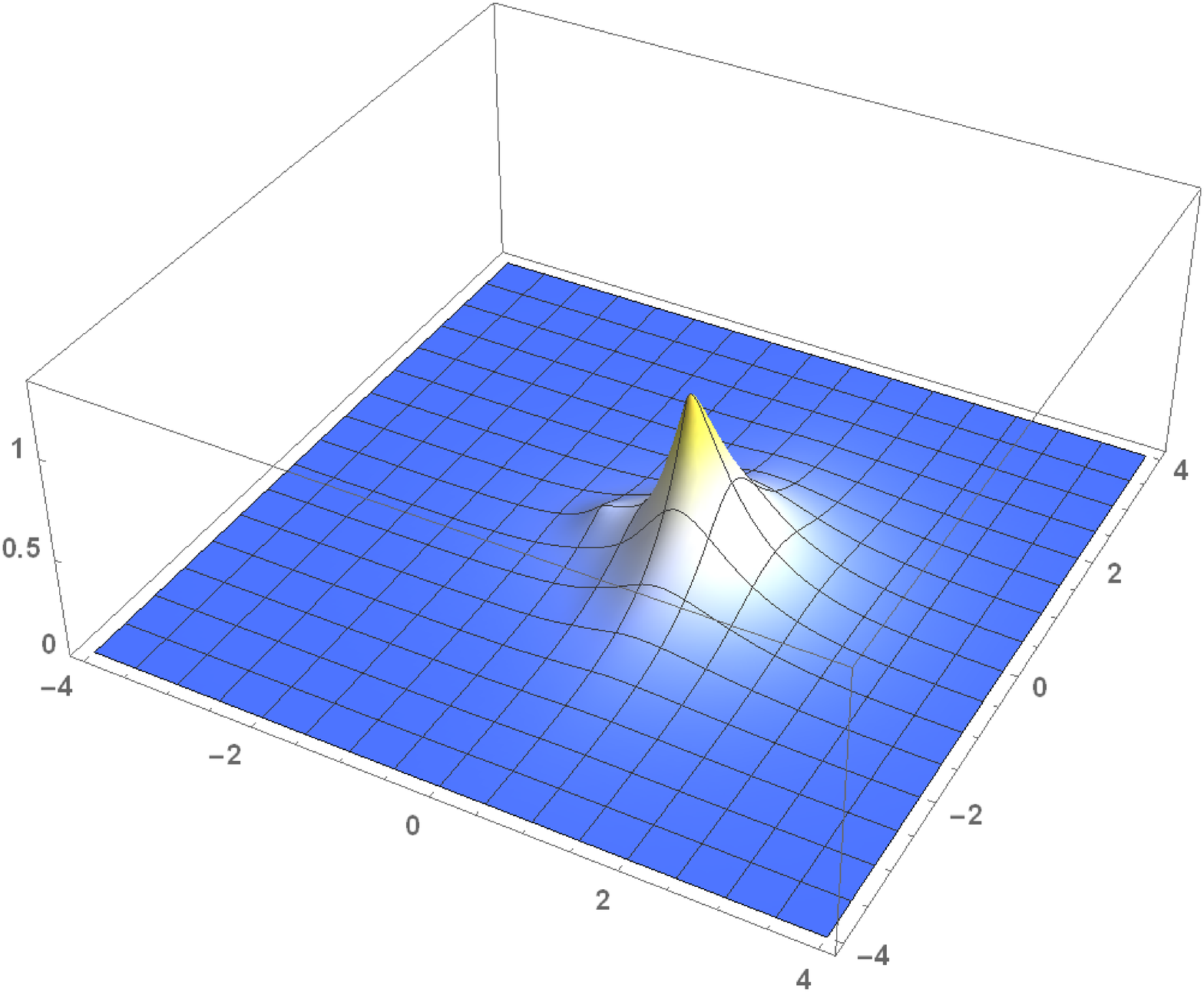}
  \includegraphics[scale=0.12]{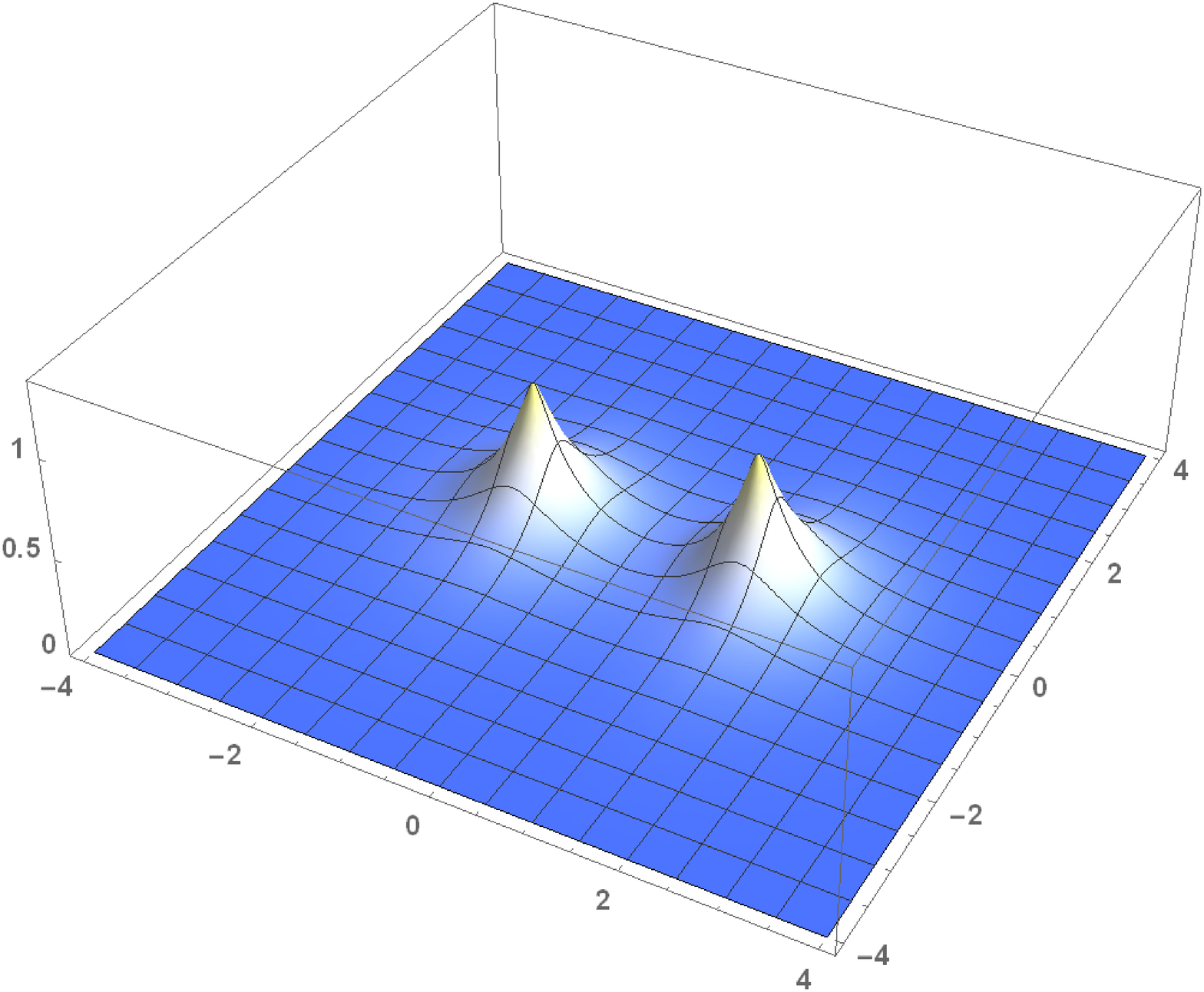}
  \includegraphics[scale=0.12]{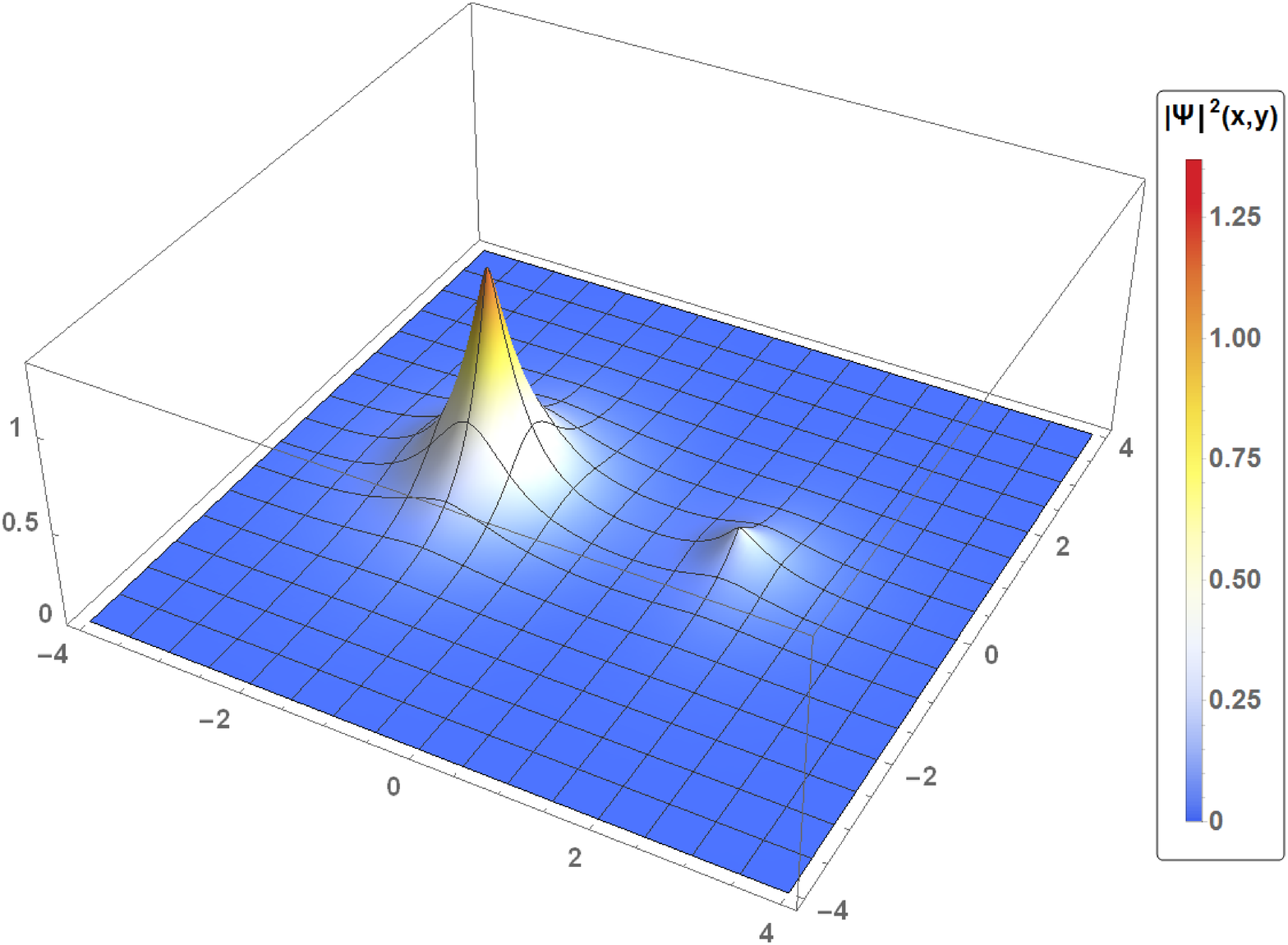}
  \caption{(Color online) The square modulus of the wave function of the lower energy level for $\zeta=0.85$ and various distances between the  
  impurities: $R=1.0$ (left panel), $R=2.55$ (middle panel), and $R=3.0$ (right panel). The wave function changes its localization from the 
  negatively charged impurity to the positively charged one as the distance between the impurities increases.}
  \label{Var_yes_wf}
\end{figure}

The approximate spectrum of the Dirac equation for the electron in the electric dipole potential with $r_{0}=0.05 R_{\Delta}$ is plotted in
Fig. \ref{sp-variational} for $\zeta=0.6$ (green dashed lines) and $\zeta=0.85$ (blue solid lines). For $\zeta=0.6$, the energy of the lowest 
electron bound state in the single positive Coulomb center problem (see Fig.\ref{fig1-LCAO}) is positive. Like in the 1D model, we 
find then that the wave function doesn't change its localization on the impurities (see Fig.~\ref{Var_no_wf}). For $\zeta=0.85$, the energy of 
the lowest electron bound state in the single
positive Coulomb center problem (see Fig.~\ref{fig1-LCAO}) is negative (for the chosen charge $\zeta=0.85$, the single positive Coulomb center 
has only one negative energy level). Therefore, the wave function changes its localization on the charged impurities (see 
Fig.~\ref{Var_yes_wf}). Our analysis performed by making use of the GK variational method shows that the migration of the wave function of the 
highest energy occupied electron bound state takes place when the charges of impurities are such that the energy levels 
of the corresponding single Coulomb centers with charges $\pm Q$ cross.

For larger $\zeta$, more complex behavior of the energy levels compared to that given by the blue solid
lines in Fig.\ref{sp-variational} with more cycles of oscillations of energy levels is observed as found in Ref.~[\onlinecite{Egger}] and 
qualitatively similar to that in Fig.\ref{second_osc} in the exactly solvable 1D model in Appendix \ref{2nd-in-1Dmodel}. It is shown
in Appendix \ref{2nd-in-1Dmodel} that the oscillations of energy levels in the 1D model correspond to the oscillations of the localization of 
the wave function from the potential barrier to well.

\section{Asymmetric case}
\label{asymmetric}

By using the variational Galerkin--Kantorovich method, we established in the previous section the supercritical instability for quasiparticles 
in graphene in the electric dipole potential. This instability is connected with the change of the localization of the wave function of the 
highest energy occupied electron bound state. We saw  that the necessary condition for the supercriticality to occur is that the energy of the 
electron bound state in the field of the single positively charged impurity cross zero. Since there is a charge conjugated bound state level in 
the single negatively charged impurity problem, taken together these two levels traverse the energy distance $2\Delta$. Recall that the 
supercritical instability in the single Coulomb center problem takes place when the lowest energy electron bound state traverses also the 
distance $2\Delta$. This suggests that if we consider the Dirac equation for quasiparticles in graphene with two impurities whose charges have 
opposite sign, however, are not equal by modulus, the supercritical instability will take place only if the lowest and highest bound state 
levels of the corresponding single positively and negatively charged impurity problems traverse together the energy distance $2\Delta$. We will 
study this suggestion in this section by considering the Dirac equation for quasiparticles in graphene with two impurities whose charges are 
opposite in sign and not equal by modulus. It is obvious that the intrinsic particle-hole symmetry defined by the operator $\Omega$ in
Sec.\ref{section-equation} is no longer present in the case of two oppositely charged impurities with charges not equal by modulus.

The Hamiltonian of this problem is given by Eq.(\ref{general-Hamiltonian}) with the potential
\begin{equation}
V(\mathbf{r})=e\left(-\frac{Z_{1}}{\sqrt{(x-R/2)^{2}+y^{2}+r_{0}^{2}}}+\frac{Z_{2}}{\sqrt{(x+R/2)^{2}
+y^{2}+r_{0}^{2}}}\right).
\label{potential-asymmetric}
\end{equation}
In terms of dimensionless quantities, the Dirac equation for the electrons in the $K_+$ valley has form (\ref{system-equations}). The Dirac 
Hamiltonian with potential (\ref{potential-asymmetric}) has the
discrete symmetry described by the operator $U$ introduced in Sec.\ref{section-equation}. The components
of the wave function $|\Psi_{+}\rangle$ satisfy conditions (\ref{components}). In order to solve the
Dirac equation with potential (\ref{potential-asymmetric}), we use the variational Galerkin--Kantorovich method and utilize the trial wave 
functions (\ref{system2}) which satisfy the asymptotic at large distance and conditions (\ref{components}).

\begin{figure}[ht]
  \centering
  \includegraphics[scale=0.35]{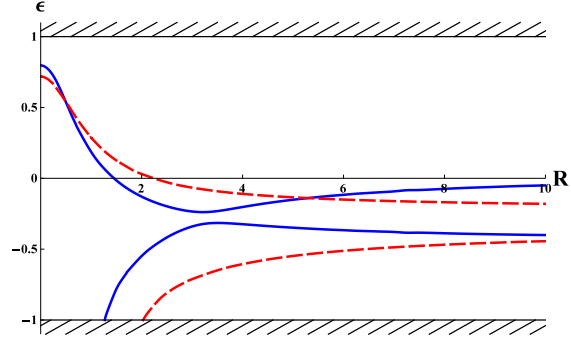}
  \caption{(Color online) The energy levels in the problem of two oppositely charged impurities with $\zeta_{1}=0.5$, $\zeta_{2}=0.8$ (red 
  dashed lines) and $\zeta_{1}=0.7$, $\zeta_{2}=0.95$ (blue solid lines) for $r_{0}=0.05 R_{\Delta}$ found in the variational method. }
  \label{fig1-asymmetric}
\end{figure}

\begin{figure}[ht]
  \centering
  \includegraphics[scale=0.2]{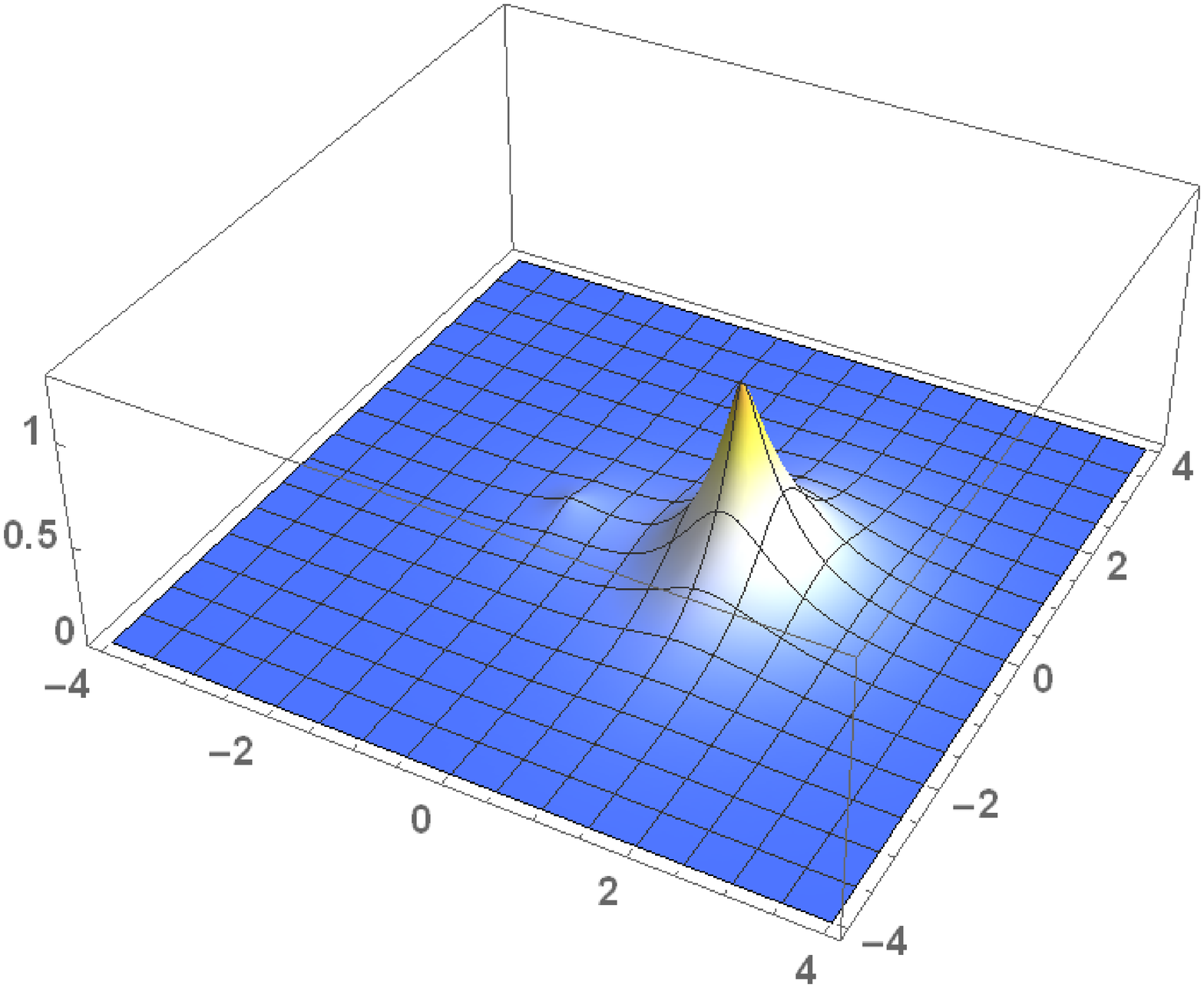}
  \includegraphics[scale=0.2]{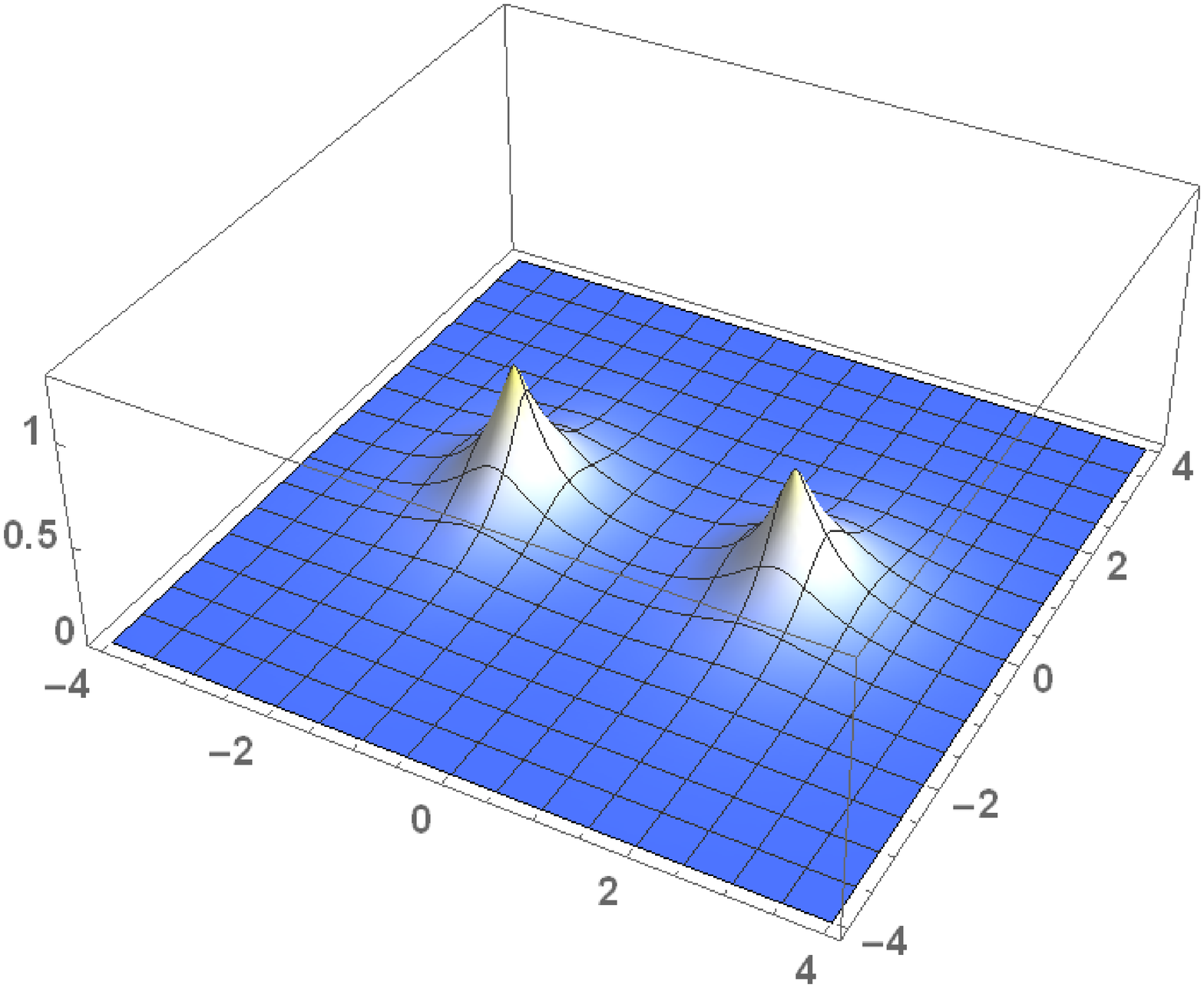}
  \includegraphics[scale=0.2]{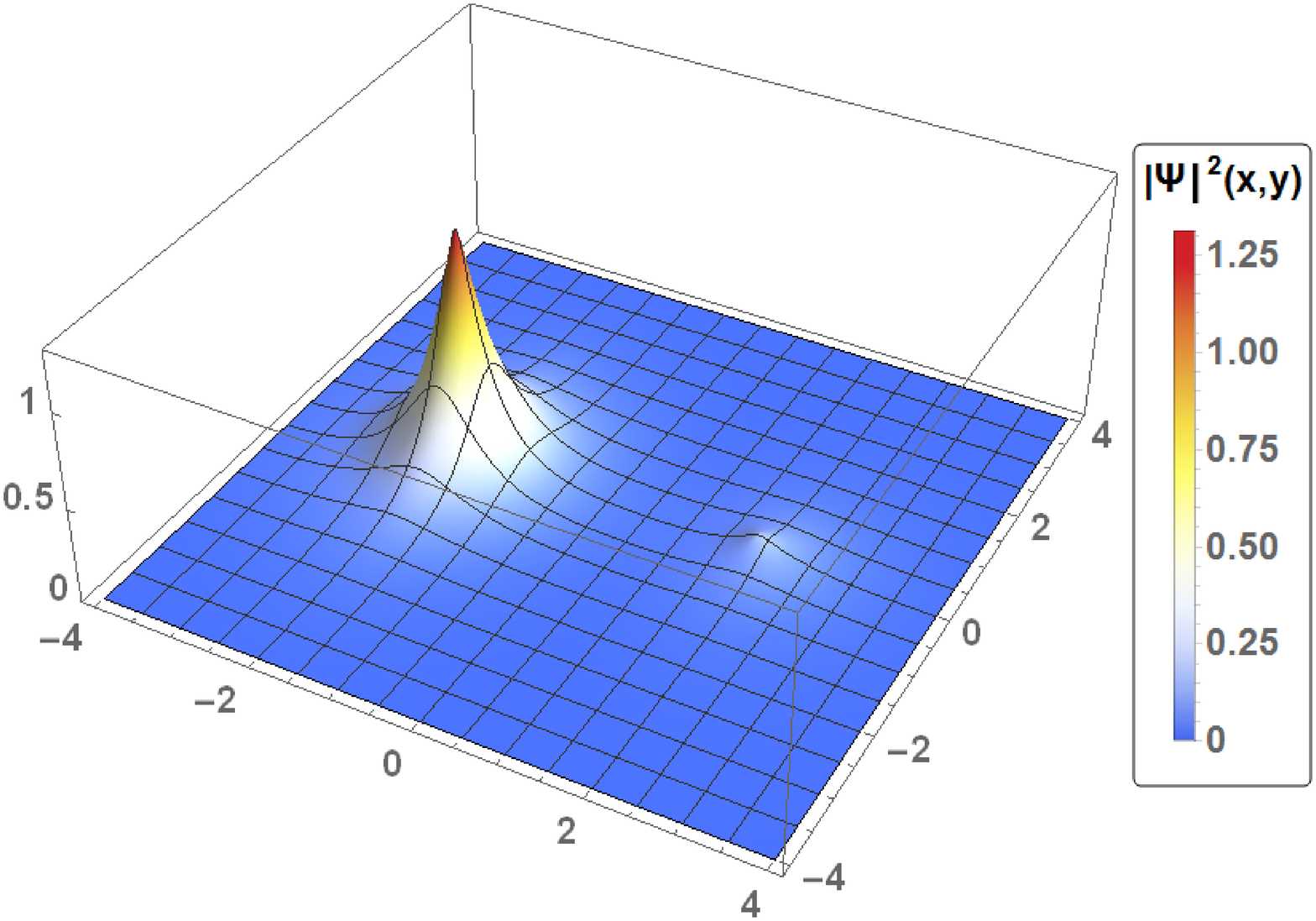}
  \caption{(Color online) The square modulus of the wave function of the lower energy level 
  for $\zeta_{1}=0.7$, $\zeta_{2}=0.95$ and various distances between impurities: $R=2.0$ (left panel), $R=3.25$ (middle panel), and $R=3.75$ 
  (right panel). The wave function changes its localization from the negatively charged impurity to the positively charged one as the distance 
  between the impurities increases.}
  \label{Asym_yes_wf}
\end{figure}

According to the Galerkin--Kantorovich method, we substitute the trial wave functions into the Dirac equation and require their orthogonality to 
the residual with respect to the variable $y$. We obtain the same system of ordinary differential equations as in the symmetric case 
(\ref{variational-method}) with the coefficient functions (\ref{P-s})-(\ref{V-s}). The coefficient function (\ref{V-s}) should be calculated
for potential (\ref{potential-asymmetric}). The condition that functions $f_{k}(x)$ and $g_{k}(x)$
are finite as $x\rightarrow\pm\infty$ allows us to determine the spectrum.

In Fig.\ref{fig1-asymmetric}, we plot the bound state energy levels in two cases $\zeta_{1}=0.5$, $\zeta_{2}=0.8$ (red dashed lines) and 
$\zeta_{1}=0.7$, $\zeta_{2}=0.95$ (blue solid lines) for $r_{0}=0.05 R_{\Delta}$ found by using the variational method with $N=0,N^{'}=0$ (one 
term in the ansatz). Note that 
the energy levels are not particle-hole symmetric. As $R\to 0$, the energy levels correspond to the levels 
of single positive charge impurity with $Z_2-Z_1>0$. For $R\to\infty$, the energy levels tend to those related with single positive and negative 
charge impurities. The negative energy levels arise from the lower continuum at some critical distance $R_c$, and this happens only when two 
impurities become sufficiently separated.

In the case $\zeta_{1}=0.5$ and $\zeta_{2}=0.8$, the bound state energy levels monotonously depend on the distance between impurities and no 
level repulsion is observed. We checked that the wave function does not change its localization similar to the case of two oppositely charged 
impurities whose charges are
equal by modulus. For larger values $\zeta_{1}=0.7$ and $\zeta_{2}=0.95$, the energies of the bound state energy levels given by the blue solid 
lines in Fig. \ref{fig1-asymmetric} first converge and then
go away from each other. This suggests that the wave function of the lower energy bound state should change its localization. In
Fig. \ref{Asym_yes_wf}, we plot the square modulus of the wave function for the lower energy level at three different distances between the 
impurities. Clearly, the wave function of the highest energy occupied state does change its localization from the negatively charged impurity to 
the positively charged one as the distance between the impurities increases. Thus, the study of an asymmetric case in the present section 
confirms the universality of the phenomenon of the change of localization of the wave function and the necessary condition for the supercritical 
instability to occur is that the lowest and highest energy bound states of the corresponding single Coulomb center problems
traverse together the energy distance $2\Delta$.

\section{Conclusion}
\label{section-conclusion}

Motivated by a recent study of the Dirac equation for quasiparticles in the electric dipole potential in graphene, we studied in the continuum 
model the supercritical instability in the electric dipole problem
in gapped graphene. Since the variables for the Dirac equation with the electric dipole potential are not separable in any known orthogonal 
coordinate system, we used the numerical variational Galerkin--Kantorovich method with trial functions which have the correct asymptotic at 
large distances. For sufficiently large charges of impurities such that the lowest energy electron bound state of the single Coulomb center
problem crosses the zero energy $E=0$, the positive and negative energy bound state levels for the Dirac
equation in the electric dipole potential at first converge and then go away (since the levels have the
same quantum numbers, they do not cross due to the avoided crossing theorem) tending to the energy levels of the corresponding single Coulomb 
center problems. We found that the wave function of the highest energy occupied bound state level changes its localization from the negatively 
charged impurity to the positively charged one at the point of the closest convergence of the bound state levels. This migration of the wave
function corresponds to a supercriticality of novel type with the spontaneous creation of an electron-hole
pair in bound electron and hole states screening the negatively and positively charged impurities, respectively.

We extended our analysis of the electric dipole problem in graphene to an asymmetric case, where the charges of impurities are of opposite sign 
and not equal by modulus. We found in this case that the wave function of the highest energy occupied bound state changes its localization only 
if the bound state levels in the corresponding single Coulomb center problems traverse {\it together} the energy distance $2\Delta$
separating the upper and lower continua.
The supercritical instability of the electric dipole can be observed experimentally by placing subcritical oppositely charged impurities on 
graphene and then moving with the tip of scanning tunneling microscope
one impurity toward the other one and afterwards moving the impurities apart again. The supercritical instability takes place if the impurities 
become screened.

Since the LCAO and variational Galerkin--Kantorovich methods are approximate ones, in order to study the robustness and validity of the 
supercriticality of novel type we studied an exactly solvable 1D model of the Dirac equation with a square potential well and 
barrier modeling an electric dipole potential. Our findings in this exactly solvable 1D model unambiguously demonstrate the presence of the 
supercriticality of novel type connected with the migration of the wave function of the  electron bound state.

\begin{acknowledgments}
We are grateful to V.M. Loktev for useful discussions. This work is supported partially by the Program of Fundamental Research of the Physics 
and Astronomy Division of the NAS of Ukraine.
\end{acknowledgments}

\appendix
\section{Oscillations of energy levels in $1D$ model with an electric-dipole-like potential}
\label{2nd-in-1Dmodel}

In the $1D$ Dirac model (\ref{hamiltonian-1D}) with the electric-dipole-like potential (\ref{dipole-1D}), the energy spectrum is determined from 
the secular equation $\det A=0$, where the matrix elements $A_{ik}$ 
are given by
\begin{eqnarray}
A_{11}&=&-A_{78}=e^{-(a+d)\kappa_{0}},\ \ \ A_{12}=\sin\left[(a+d)k_{1}\right], \ \ \ A_{13}=-\cos\left[(a+d)k_{1}\right],\nonumber\\
A_{21}&=&A_{88}=\frac{\kappa_{0}e^{-(a+d)\kappa_{0}}}{1+\epsilon},\ \ \ A_{22}=-\frac{k_{1}\cos\left[(a+d)k_{1}\right]}{1+\epsilon+v_{0}},\ \ \
A_{23}=-\frac{k_{1}\sin\left[(a+d)k_{1}\right]}{1+\epsilon+v_{0}},\nonumber\\
A_{32}&=&-\sin\left[(a-d)k_{1}\right], \ \ \ A_{33}=\cos\left[(a-d)k_{1}\right], \ \ \ A_{55}=-A_{34}=e^{-(a-d)\kappa_{0}},\nonumber\\
A_{54}&=&-A_{35}=e^{(a-d)\kappa_{0}}, \ \ \ A_{56}=-\sin\left[(a-d)k_{2}\right], \ \ \ A_{57}=-\cos\left[(a-d)k_{2}\right],\nonumber\\
A_{42}&=&\frac{k_{1}\cos\left[(a-d)k_{1}\right]}{1+\epsilon+v_{0}}, \ \ \ A_{43}=\frac{k_{1}\sin\left[(a-d)k_{1}\right]}{1+\epsilon+v_{0}},
\ \ \ A_{44}=A_{65}=-\frac{\kappa_{0}e^{-(a-d)\kappa_{0}}}{1+\epsilon},\nonumber\\
A_{45}&=&A_{64}=\frac{\kappa_{0}e^{(a-d)\kappa_{0}}}{1+\epsilon}, \ \ \ A_{66}=-\frac{k_{2}\cos\left[(a-d)k_{2}\right]}{1+\epsilon-v_{0}}, \ \ \
A_{67}=\frac{k_{2}\sin\left[(a-d)k_{2}\right]}{1+\epsilon-v_{0}},\nonumber\\
A_{76}&=&\sin\left[(a+d)k_{2}\right], \ \ \ A_{77}=\cos\left[(a+d)k_{2}\right],\nonumber\\
A_{86}&=&\frac{k_{2}\cos\left[(a+d)k_{2}\right]}{1+\epsilon-v_{0}}, \ \ \ A_{87}=-\frac{k_{2}\sin\left[(a+d)k_{2}\right]}{1+\epsilon-v_{0}},
\label{Aij-coeff}
\end{eqnarray}
and the rest of elements equals zero.

\begin{figure}[ht]
  \centering
  \includegraphics[scale=0.35]{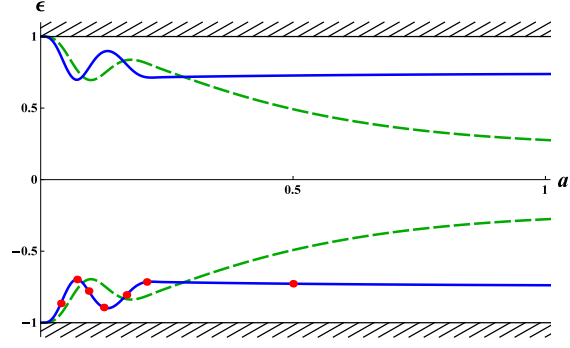}
  \caption{(Color online) The bound state energy levels of the 1D problem for $v_{0}=9$ (green dashed lines) and $v_{0}=12$ (blue solid lines) 
  obtained for $d=0.25$.}
  \label{second_osc}
\end{figure}

We analysed in Sec.\ref{A} the behavior of energy levels for sufficiently small value $v_0$. It is instructive to consider what happens for 
larger values of $v_0$. We plot in Fig.~\ref{second_osc} the
bound state energy levels in the ``electric-dipole-like'' potential for $d=0.25$ and two values of the strength of the potential: $v_{0}=9$ 
(green dashed lines) and $v_{0}=12$ (blue solid lines). According
to Fig.~\ref{1well-energy}, for $v_{0}=9$, only one energy level in the potential well problem crosses
the zero energy $\epsilon=0$. This corresponds to the only point of level repulsion (the point of maximal convergence of levels) in 
Fig.~\ref{second_osc} (green dashed lines). For $v_{0}=12$, two levels in the potential well cross the zero energy $\epsilon=0$ (see 
Fig.~\ref{1well-energy}). This corresponds to 
the two points of level repulsion in Fig.~\ref{second_osc} (blue solid lines) and between these two points there is a point of maximal 
divergence of energy levels. Thus, a qualitatively new phenomenon connected with oscillations of energy levels appears for sufficiently large 
values of $v_0$.

In order to demonstrate how the oscillations of energy levels for $v_0=12$ are reflected in the localization property of the wave function of 
the highest energy bound state, we plot in Fig.~\ref{fig8-toy} the square modulus of the electron wave function {of the negative energy bound 
state at seven values of $a$ shown by red points in Fig.\ref{second_osc}. One can see in Fig.\ref{fig8-toy} (b) that the electron wave function 
is localized both on the barrier and well at the first point of the maximal convergence of the energy levels. According to
Fig.\ref{fig8-toy}~(d), the wave function at the point of the maximal divergence of levels between the points of the maximal convergence is 
localized mainly on the well. At the second point of the maximal convergence of the energy levels, the wave function whose square modulus in 
plotted in Fig.\ref{fig8-toy}~(f) is again localized both on the well and barrier. As the distance between the centers $2a$ increases further, 
the electron wave function becomes localized on the well.

For still larger values of $v_0$, we observe more oscillations of energy levels which qualitatively resemble those found in 
Ref.~[\onlinecite{Egger}] for the electric dipole problem in graphene.

\begin{figure}[ht]
  \centering
  \includegraphics[scale=0.4]{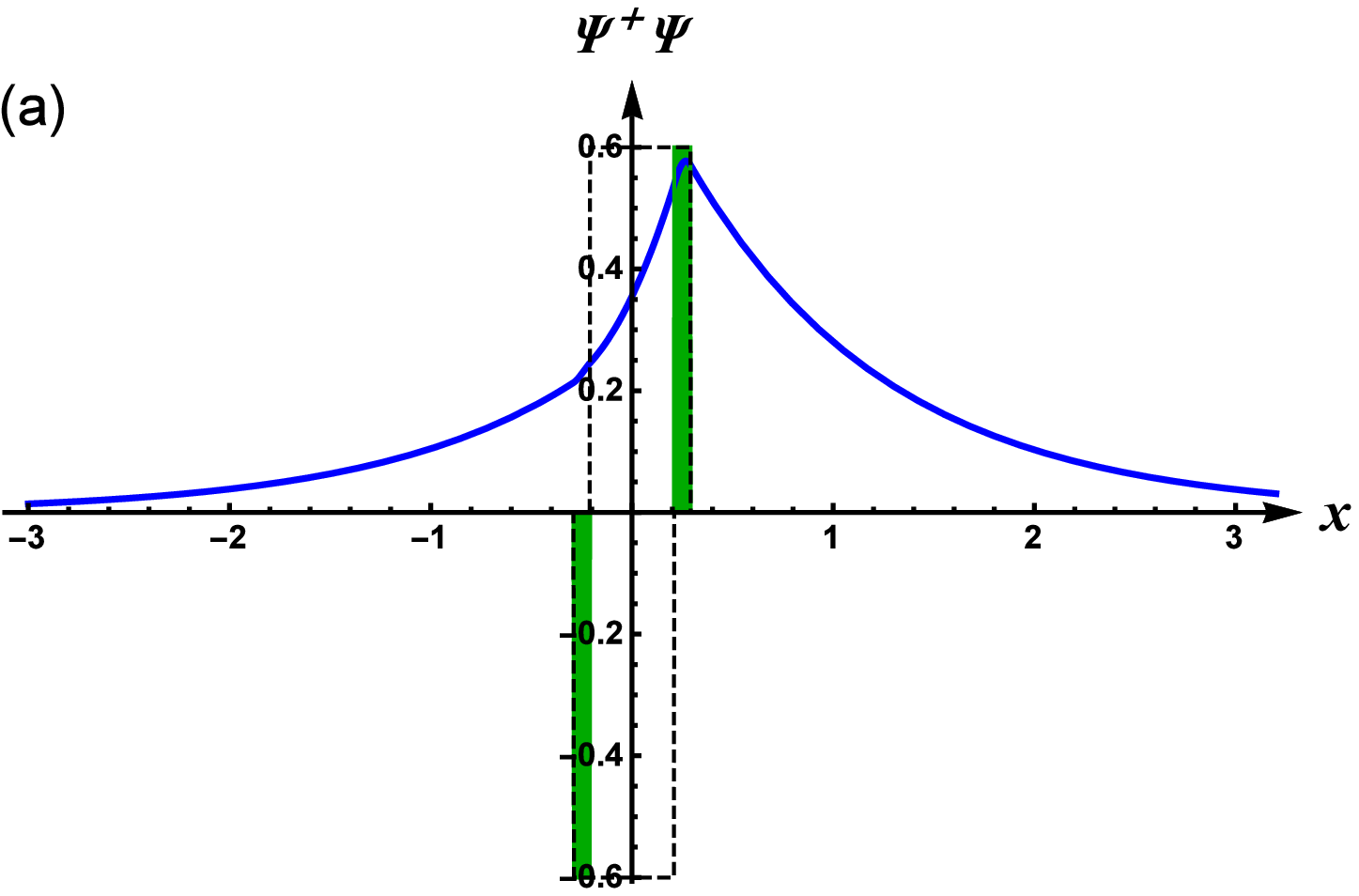}
  \includegraphics[scale=0.4]{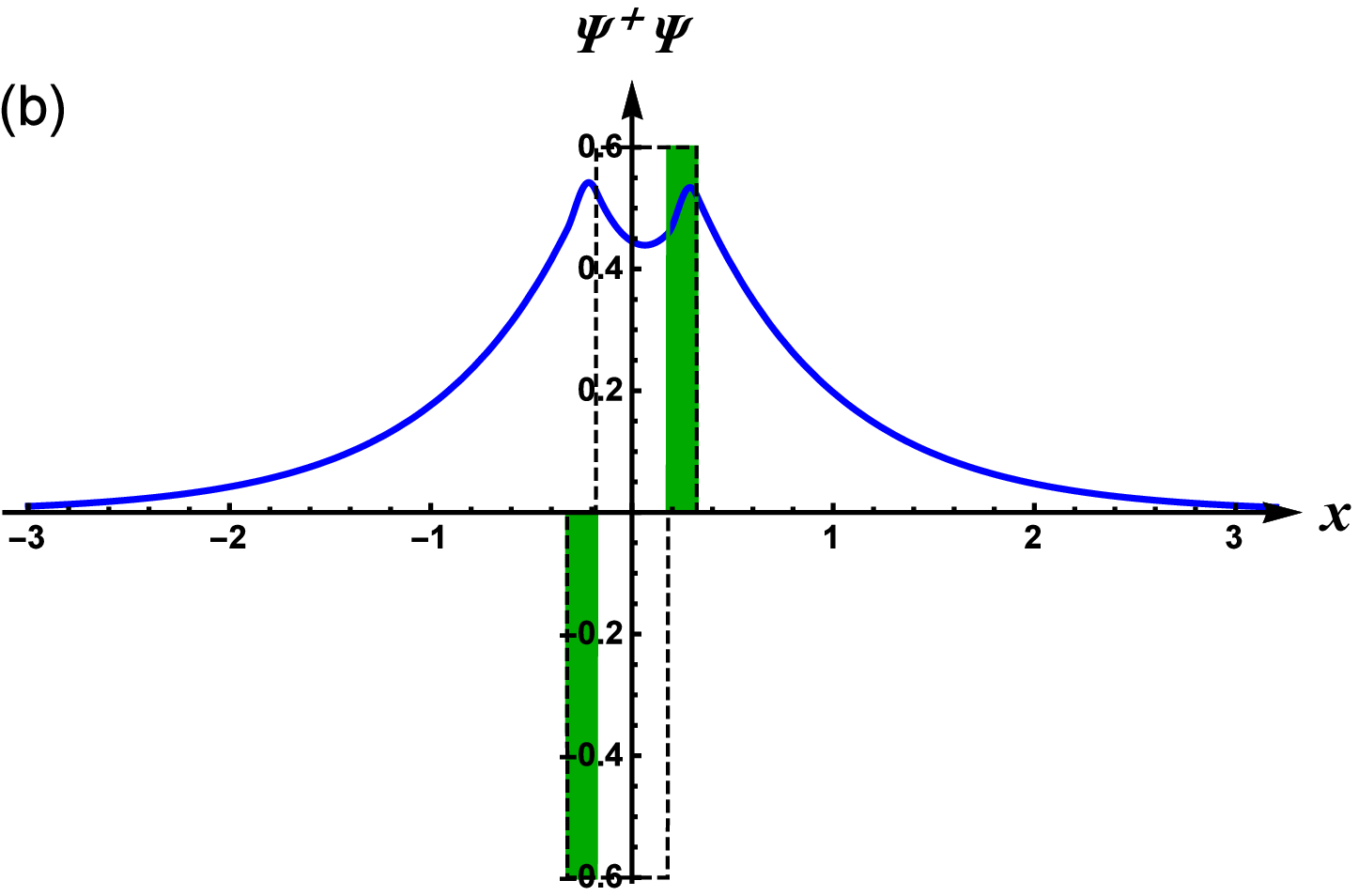}
  \includegraphics[scale=0.4]{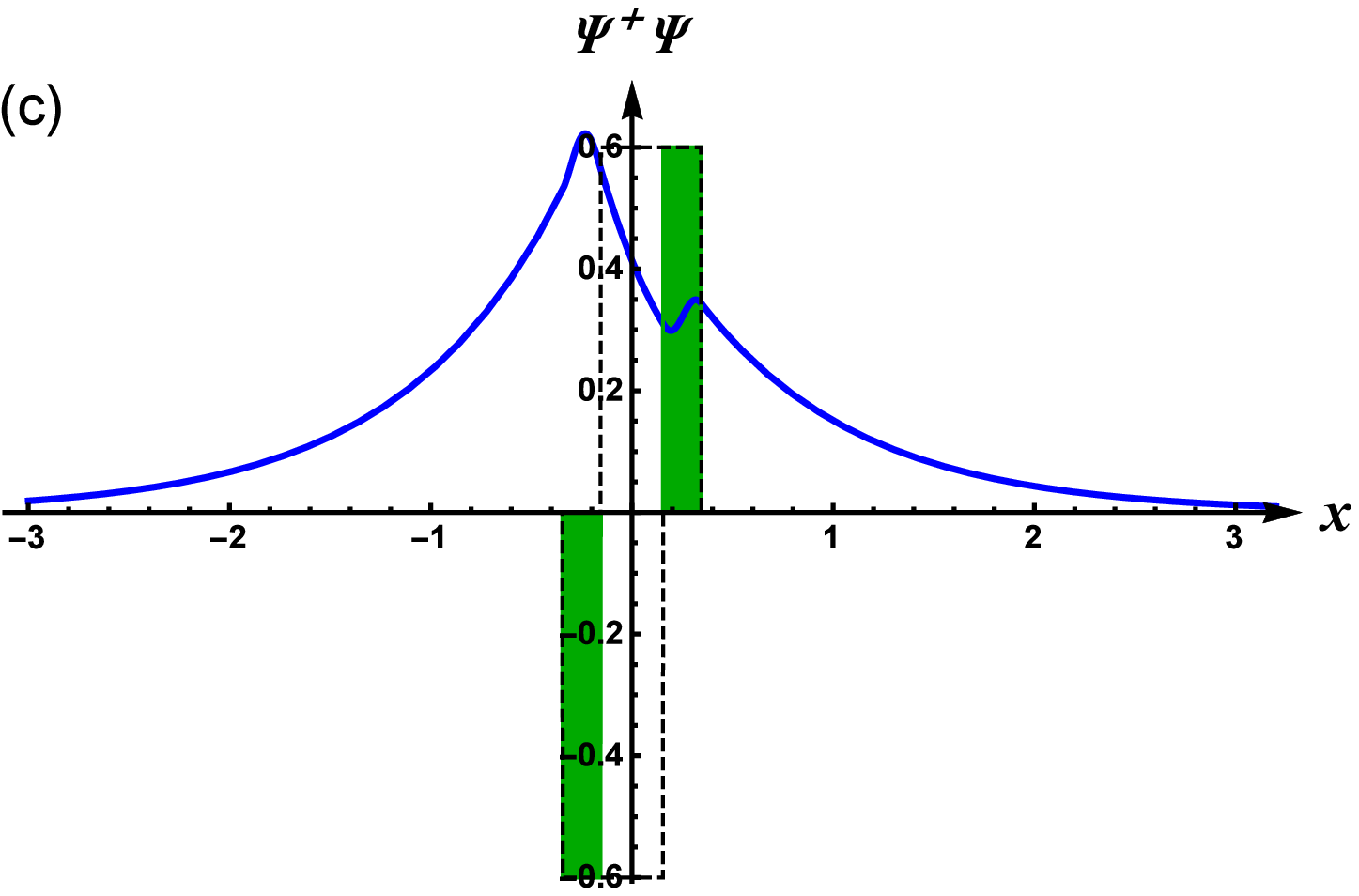}
  \includegraphics[scale=0.4]{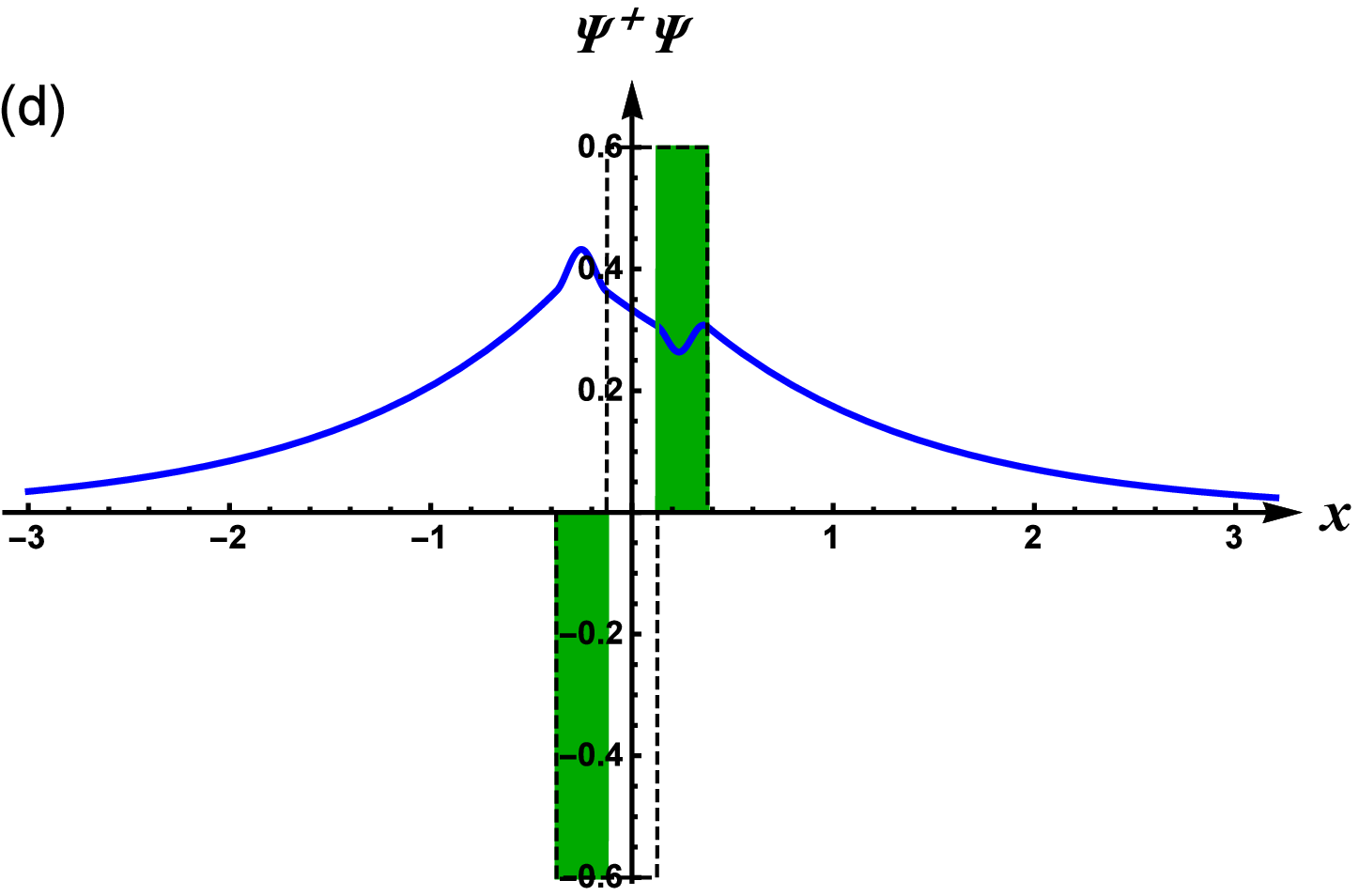}
  \includegraphics[scale=0.4]{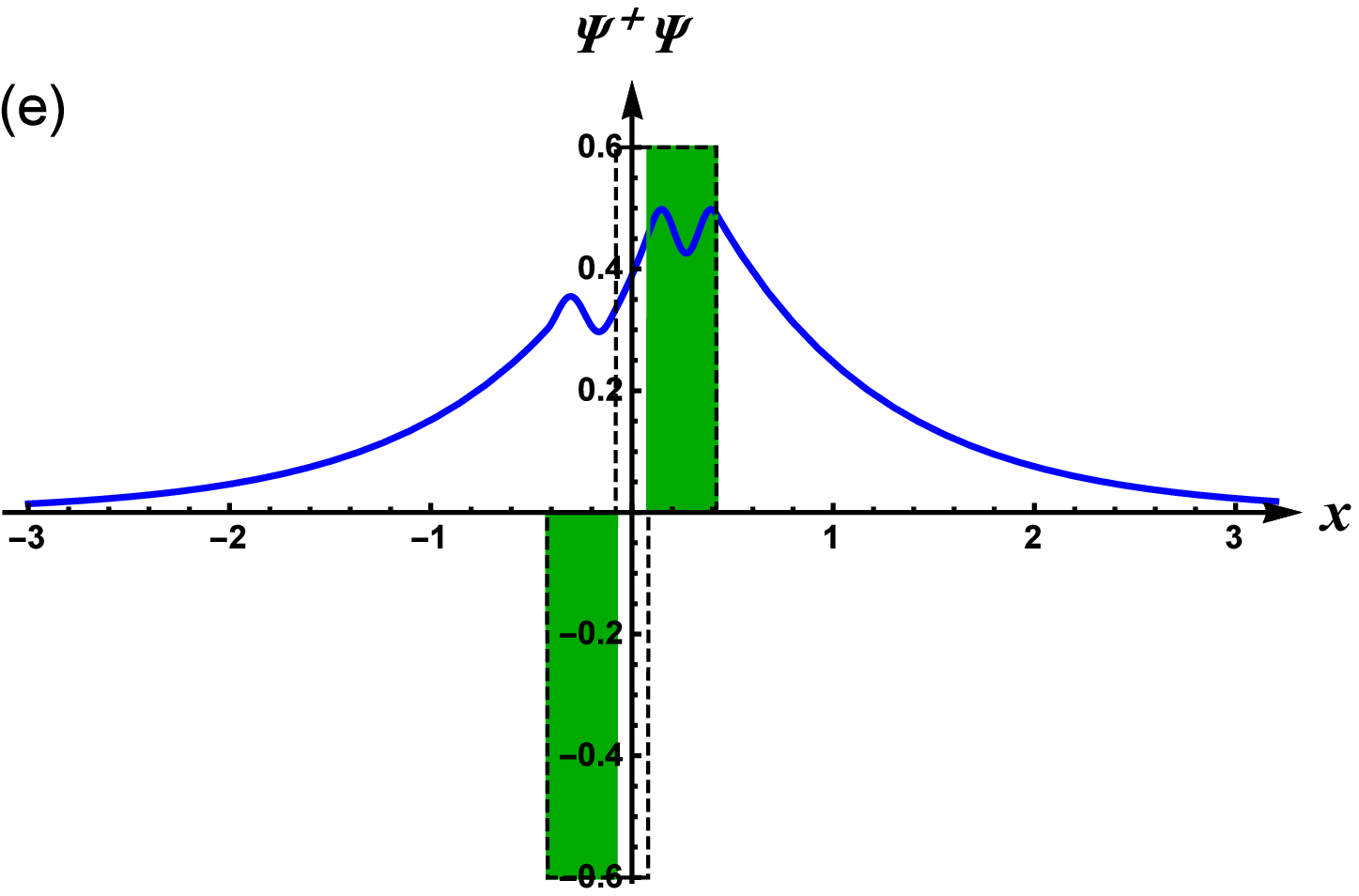}
  \includegraphics[scale=0.4]{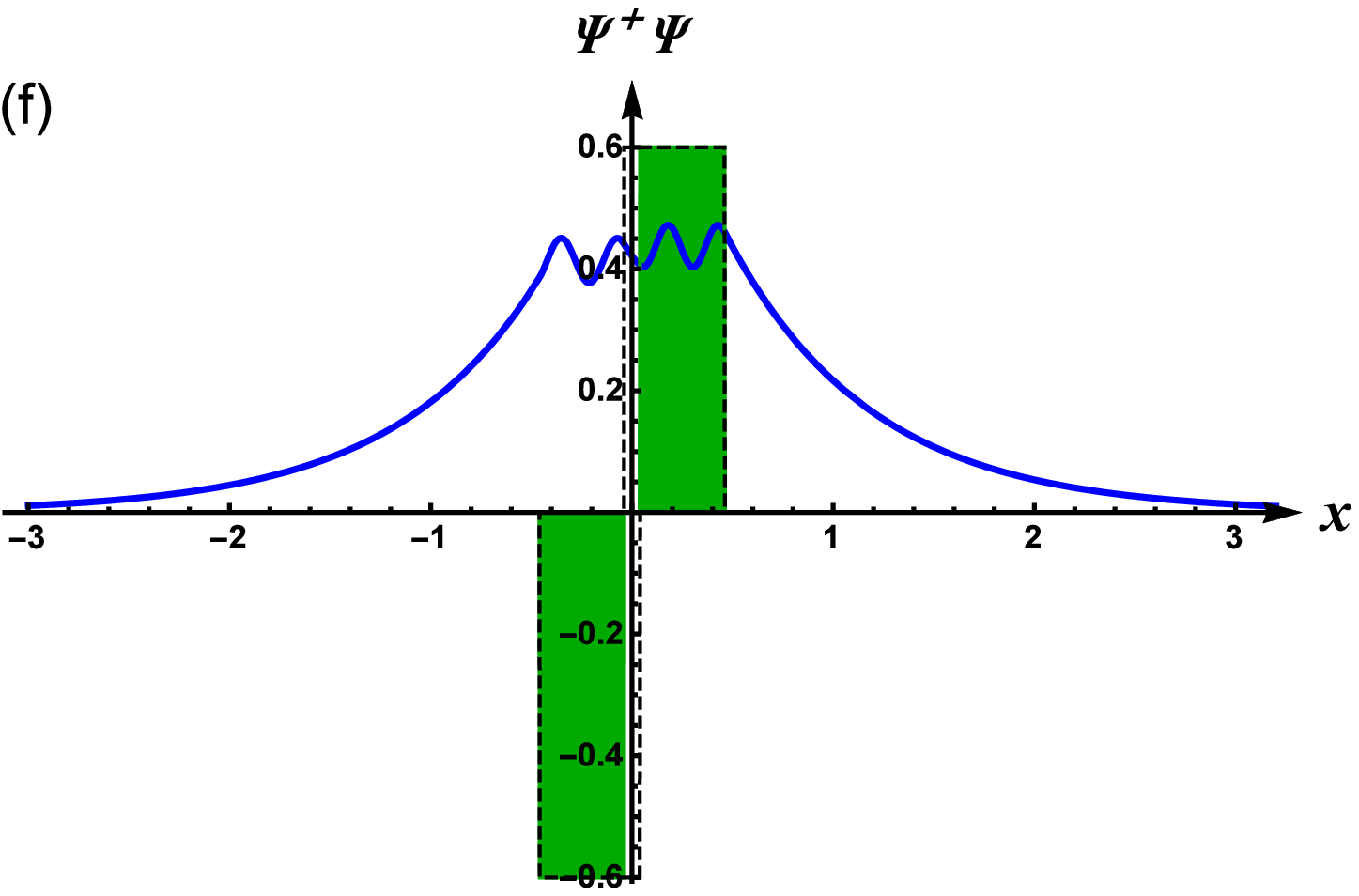}
  \includegraphics[scale=0.4]{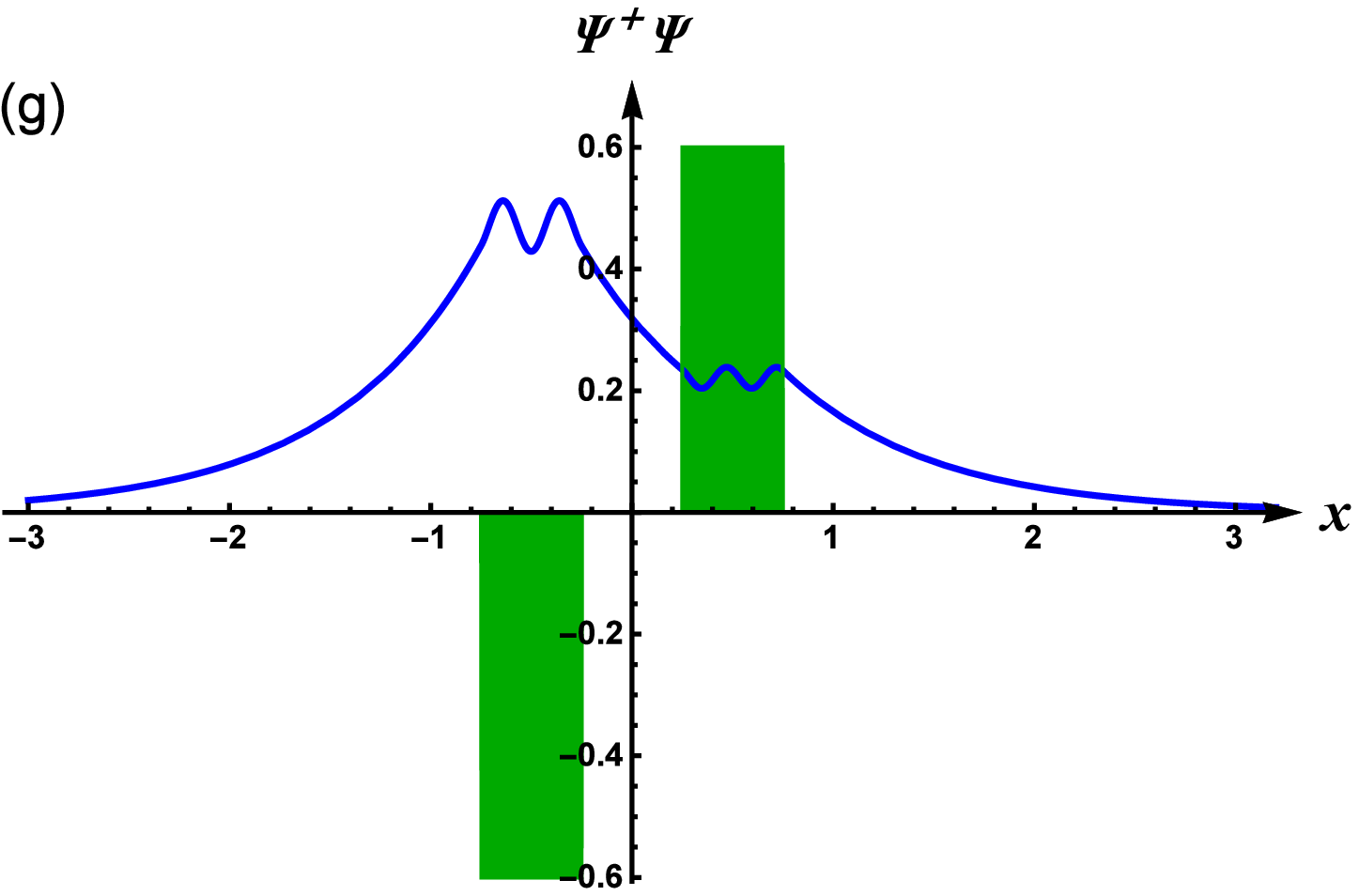}
  \caption{(Color online) The square modulus of the wave function of the negative energy bound state for $v_{0}=12$, $d=0.25$, and seven values 
  of the distance between the centers of the well and barrier: (a) $2a=0.08$, (b) $2a=0.144$, (c) $2a=0.19$, (d) $2a=0.25$, (e) $2a=0.34$, (f) 
  $2a=0.42$, (g) $2a=1.0$. 
  The corresponding values of $a$ are marked by red points in Fig.~\ref{second_osc}. The potentials of 
  the well and barrier are schematically plotted as filled green regions.}
  \label{fig8-toy}
\end{figure}

\section{Differentional equations in the Galerkin--Kantorovich variational method}
\label{B}

For the trial functions $f_k$ and $g_k$ in Eq.(\ref{system2}), we find the following system of equations
$(l=\overline{0,N};\ \ l^{'}=\overline{1,N^{'}})$:
\begin{eqnarray}
&&\sum\limits_{k=0}^{N}\left\{P_{k+l}\left(\frac{df_{2k}(x)}{dx}-(1+\epsilon)g_{2k}(x)\right)
-\sqrt{1-\epsilon^{2}}
\left(x-\frac{R}{2}{\rm sign}(x)\right)Q_{k+l}f_{2k}(x)+V_{k+l}g_{2k}(x)\right\}-\nonumber\\
&&-\sum\limits_{k=1}^{N^{'}}\left\{(2k-1)P_{k+l-1}-\sqrt{1-\epsilon^{2}}Q_{k+l}
\right\}f_{2k-1}(x)=0,\nonumber\\
&&\sum\limits_{k=1}^{N^{'}}\left\{P_{k+l^{'}-1}\left(\frac{df_{2k-1}(x)}{dx}
-(1+\epsilon)g_{2k-1}(x)\right)-\sqrt{1-\epsilon^{2}}\left(x-\frac{R}{2}{\rm sign}(x)\right)Q_{k+l^{'}-1}f_{2k-1}(x)
+V_{k+l^{'}-1}g_{2k-1}(x)\right\}+\nonumber\\
&&+\sum\limits_{k=0}^{N}\left\{2k P_{k+l^{'}-1}-\sqrt{1-\epsilon^{2}}Q_{k+l^{'}}\right\}f_{2k}(x)=0,
\nonumber\\
&&\sum\limits_{k=0}^{N}\left\{P_{k+l}\left(\frac{dg_{2k}(x)}{dx}-(1-\epsilon)f_{2k}(x)\right)
-\sqrt{1-\epsilon^{2}}
\left(x-\frac{R}{2}{\rm sign}(x)\right)Q_{k+l}g_{2k}(x)-V_{k+l}f_{2k}(x)\right\}+\nonumber\\
&&+\sum\limits_{k=1}^{N^{'}}\left\{(2k-1)P_{k+l-1}-\sqrt{1-\epsilon^{2}}Q_{k+l}
\right\}g_{2k-1}(x)=0,\nonumber\\
&&\sum\limits_{k=1}^{N^{'}}\left\{P_{k+l^{'}-1}\left(\frac{dg_{2k-1}(x)}{dx}
-(1-\epsilon)f_{2k-1}(x)\right)-\sqrt{1-\epsilon^{2}}\left(x-\frac{R}{2}{\rm sign}(x)\right)Q_{k+l^{'}-1}g_{2k-1}(x)
-V_{k+l^{'}-1}f_{2k-1}(x)\right\}-\nonumber\\
&&-\sum\limits_{k=0}^{N}\left\{2k P_{k+l^{'}-1}-\sqrt{1-\epsilon^{2}}Q_{k+l^{'}}\right\}g_{2k}(x)=0,
\label{variational-method}
\end{eqnarray}
where the coefficient functions equal
\begin{equation}
P_{s}(x)=\int\limits_{0}^{+\infty}e^{-2\sqrt{1-\epsilon^{2}}\sqrt{\left(|x|-R/2\right)^{2}+y^{2}+r_{0}^{2}}}y^{2s}dy
=\frac{\tilde{x}^{s+1}(2s-1)!!}{\alpha^{s}}{\rm K}_{s+1}(\alpha\tilde{x}),
\label{P-s}
\end{equation}
\begin{equation}
Q_{s}(x)=\int\limits_{0}^{+\infty}e^{-2\sqrt{1-\epsilon^{2}}\sqrt{\left(|x|-R/2\right)^{2}+y^{2}
+r_{0}^{2}}}\frac{y^{2s}}{\sqrt{\left(|x|-R/2\right)^{2}+y^{2}+r_{0}^{2}}}dy=
\frac{\tilde{x}^{s}(2s-1)!!}{\alpha^{s}}{\rm K}_{s}(\alpha\tilde{x}),
\label{Q-s}
\end{equation}
\begin{equation}
V_{s}(x)=\int\limits_{0}^{+\infty}e^{-2\sqrt{1-\epsilon^{2}}\sqrt{\left(|x|-R/2\right)^{2}+y^{2}
+r_{0}^{2}}}\ v(x,y)\ y^{2s}dy.
\label{V-s}
\end{equation}
Here $v(x,y)=\zeta\left(\frac{1}{\sqrt{\left(x-R/2\right)^{2}+y^{2}+r_{0}^{2}}}-\frac{1}{\sqrt{\left(x+R/2\right)^{2}+y^{2}+r_{0}^{2}}}\right)$, 
$\alpha=2 \sqrt{1-\epsilon ^2}$, $\tilde{x}=\sqrt{\left(\left| x\right| -\frac{R}{2}\right)^2+r_{0}^2}$, and ${\rm K}_s(x)$ is the MacDonald 
function. Unfortunately, function (\ref{V-s}) cannot be expressed in terms of elementary or special functions. However, it can be represented as 
a series

\begin{equation}
V_{s}(x)=\zeta {\rm sign}(x)\  \tilde{x}^{2s}\int\limits_{1}^{+\infty}e^{-z r}(r^2-1)^{s-1/2}
\left(1-\frac{r}{\sqrt{r^2+a^2}}\right)dr=\zeta {\rm sign}(x)\  \tilde{x}^{2s} \sum\limits_{n=1}^{\infty}(-1)^{n+1}
\frac{(2n-1)!!}{2^{n}n!}a^{2n}I_{s,n}(z),
\end{equation}
\begin{equation}
I_{s,n}(z)=\int\limits_{1}^{+\infty}e^{-z r}(r^2-1)^{s-1/2}\frac{dr}{r^{2n}}=\frac{(2s-1)!!}{2^{s+1}}G^{30}_{13}\left(\frac{z^{2}}{4}\left|
\begin{array}{c}n+1/2\\ \frac{1}{2},\ 0,\ (n-s)\end{array}\right.\right),
\label{integral-Meijer}
\end{equation}
where  $z=\alpha\tilde{x}$, $a^2=2|x|R/\tilde{x}^2$, and $G^{30}_{13}$ is the Meijer G-function. When evaluating integral
(\ref{integral-Meijer}) we used the formula 2.1.4.20 from book [\onlinecite{Prudnikov-v3}] and the relations of symmetry and shifting for the 
Meijer's function. The condition that functions $f_{k}(x)$ and $g_{k}(x)$ are finite as $x\rightarrow\pm\infty$ allows us 
to determine the spectrum.


\begin{thebibliography}{99}

\bibitem{Pomeranchuk} I.Ya Pomeranchuk and Y.A. Smorodinsky, J. Phys. USSR {\bf 9}, 97 (1945).

\bibitem{Zeldovich} Ya.B. Zeldovich and V.N. Popov, Sov. Phys. Usp. {\bf 14}, 673 (1972).

\bibitem{Greiner} W. Greiner, B. Muller, and J. Rafelski, {\it Quantum
Electrodynamics of Strong Fields} (Springer-Verlag, Berlin, 1985).

\bibitem{Pereira} V.M. Pereira, J. Nilsson, A.H. Castro Neto, Phys. Rev. Lett. {\bf 99}, 166802 (2007).

\bibitem{Shytov} A.V. Shytov, M.I. Katsnelson, and L.S. Levitov, Phys. Rev.
Lett. {\bf 99}, 236801 (2007); {\it ibid}, {\bf 99}, 246802 (2007).

\bibitem{Novikov} M.M. Fogler, D.S. Novikov, and B.I. Shklovskii, Phys. Rev. B {\bf 76}, 233402 (2007).

\bibitem{excitonic-instability} O.V. Gamayun, E.V. Gorbar, and V.P. Gusynin,
Phys. Rev. B {\bf 80}, 165429 (2009).

\bibitem{Fertig}J.~Wang, H.A.~Fertig, and G.~Murthy, Phys. Rev. Lett. {\bf 104}, 186401 (2010).

\bibitem{Guinea}J.~Sabio, F.~Sols, and F.~Guinea, Phys. Rev. B {\bf81}, 045428 (2010).

\bibitem{metal-insulator}D.V. Khveshchenko, Phys. Rev. Lett. {\bf 87}, 246802 (2001);
E.V. Gorbar, V.P. Gusynin, V.A. Miransky, and I.A.~Shovkovy, Phys. Rev. B {\bf 66},
045108 (2002);
ibid, Phys. Lett. A {\bf313}, 472 (2003);
D.V.~Khveshchenko and H.~Leal, Nucl. Phys. B {\bf687}, 323 (2004).
\bibitem{GGG2010} O.V. Gamayun, E.V. Gorbar, and V.P. Gusynin,
Phys. Rev. B {\bf 81}, 075429 (2010).

\bibitem{MS-phase-transition}J.E.~Drut and T.A.~L$\ddot{a}$hde, Phys. Rev. Lett.
{\bf102}, 026802 (2009);
Phys. Rev. {\bf79}, 241405(R) (2009);
W.~Armour, S.~Hands, and C.~Strouthos,
Phys. Rev. B {\bf81}, 125105 (2010);
P.V.~Buividovich and M.I.~Polikarpov, Phys. Rev. B {\bf86}, 245117 (2012).

\bibitem{Gonzalez}J.~Gonz$\acute{a}$lez, Phys. Rev. B {\bf85}, 085420 (2012).

\bibitem{Wang}Y.~Wang et al., Science {\bf340}, 734 (2013).

\bibitem{Kirczenow} A. Saffarzadeh and G. Kirczenow, Phys. Rev. B {\bf 90}, 155404 (2014).

\bibitem{two-centers} O.O. Sobol, E.V. Gorbar, and V.P. Gusynin, Phys. Rev. B {\bf 88}, 205116 (2013).

\bibitem{Egger} A. De Martino, D. Klopfer, D. Matrasulov, and R. Egger, Phys. Rev. Lett. {\bf 112},
186603 (2014).

\bibitem{Matrasulov} D. Klopfer, A. De Martino, D. Matrasulov, and R. Egger, Eur. Phys. J. {\bf 87}, 187 (2014).

\bibitem{Matveev}D.U. Matrasulov, V.I. Matveev, and M.M. Musakhanov, Phys. Rev. A {\bf60}, 4140 (1999).

\bibitem{Connolly} K. Connolly and D.J. Griffiths, Am. J. Phys. {\bf 75}, 524 (2007).

\bibitem{Turner} J.E. Turner, Am. J. Phys. {\bf 45}, 758 (1977).

\bibitem{Wigner} J. von Neumann and E.P. Wigner, Z. Physik {\bf 30}, 467 (1929).

\bibitem{supercriticality} E.V. Gorbar, V.P. Gusynin, and O.O. Sobol, arXiv:1506.08379 [cond-mat.str-el],
to appear in Europhysics Letters, V.111 (2015).

\bibitem{footnote}Electric dipole potential in one dimension is usually studied for the
Schr$\ddot{o}$dinger equation (see Ref.~[\onlinecite{Connolly}] and references therein). In particular, 
for the potential well and barrier as well as the delta-function dipole it was shown that these potentials admit at least one bound state for 
any separation distances between charges. We are not aware about similar
studies for 1D Dirac equation (for the double-delta function potential in 1D Dirac equation,
see Ref.[\onlinecite{double-delta}]).

\bibitem{double-delta}F.~Fillion-Gourdeau, E.~Lorin, and A.D.~Bandrauk, J. Phys. A: Math. Theor.
{\bf45}, 215304 (2012).

\bibitem{Vasak} D. Vasak, K.H. Wietschorke, B. Muller, and W. Greiner, Z. Phys. {\bf C21}, 119 (1983).

\bibitem{animation} See Supplemental Material in video format at [URL will be inserted by publisher] for 
the dependence of the local density of states on the distance $a$ in the 1D problem with an electric-dipole-like potential.

\bibitem{Peres} A.H. Castro Neto, F. Guinea, N.M.R. Peres, K.S. Novoselov, and A.K. Geim, Rev. Mod. Phys. {\bf 81}, 109 (2009).

\bibitem{Ponomarenko} L.A. Ponomarenko et al., Nature {\bf 497}, 594 (2013).

\bibitem{Song} J.C.W. Song, A.V. Shytov, and L.S. Levitov, Phys. Rev. Lett. {\bf 111}, 266801 (2013).

\bibitem{McDonald} F. Zhang, J. Jung, G.A. Fiete, Q. Niu, and A.H. MacDonald, Phys. Rev. Lett. {\bf 106}, 156801 (2011).

\bibitem{Kotov} V.M. Pereira, V.N. Kotov, and A.H. Castro Neto, Phys. Rev. B {\bf 78}, 085101 (2008).

\bibitem{Prudnikov-v3} A.P.~Prudnikov, Yu.A.~Brychkov, and O.I.~Marichev, {\it Integrals and Series. Direct
Laplace Transforms}. Vol.4 (Gordon and Breach Sci. Publishers, New York, 1992).


\end{thebibliography}
\end{document}